\documentclass[a4paper]{article}
\usepackage{jheppub}
\usepackage{amssymb}   % for math
\usepackage{amsmath}
\usepackage{mathtools}
\usepackage{makeidx}
\usepackage{graphicx}
\usepackage{caption}
\usepackage{subcaption}
\usepackage[utf8]{inputenc}
\usepackage{feynmp-auto}
\usepackage{graphicx}
\usepackage[draft]{fixme}
\usepackage{booktabs} % for book-like tables
\def\beq{\begin{equation}}
\def\eeq{\end{equation}}
\def\bea{\begin{eqnarray}}
\def\eea{\end{eqnarray}}

\def\f21{{}_2F_{1}}

\def\t{\widetilde}
\def\m{\mathrm{m}}

\hypersetup{colorlinks = false}

\def\bsp#1\esp{\begin{split}#1\end{split}}

\preprint{CERN-TH-2018-018}

\title{Higgs Boson Production at Hadron Colliders at N$^3$LO in QCD}

\author[]{Bernhard Mistlberger}
\affiliation[]{CERN Theory Division, CH-1211, Geneva 23, Switzerland}
\abstract{
We present the Higgs boson production cross section at Hadron colliders in the gluon fusion production mode through N$^3$LO in perturbative QCD.
Specifically, we work in an effective theory where the top quark is assumed to be infinitely heavy and all other quarks are considered to be massless.
Our result is the first exact formula for a partonic hadron collider cross section at N$^3$LO in perturbative QCD.
Furthermore, this result represents the first analytic computation of a hadron collider cross section involving elliptic integrals.
We derive numerical predictions for the Higgs boson cross section at the LHC. 
Previously this result was approximated by an expansion of the cross section around the production threshold of the Higgs boson and we compare our findings.
Finally, we study the impact of our new result on the state of the art prediction for the Higgs boson cross section at the LHC.
}

\keywords{N3LO, QCD, Higgs boson}

\begin{document}
\notoc
\maketitle
\newpage
\section{Introduction}
With the discovery of the Higgs boson~\cite{Aad2012,Chatrchyan2012} at the Large Hadron Collider (LHC) at CERN we have entered a new era of particle physics phenomenology. 
With conclusive evidence for the existence of the Higgs boson the Standard Model (SM) of particle physics has become a self consistent theory.
It explains the mechanism of electro-weak symmetry breaking, the origin of elementary particle masses and it allows to derive concise predictions to energies far beyond current experimental reach.
The SM is however a phenomenologically incomplete theory and needs to be extended to obtain a satisfying description of all known physics.
Higgs boson measurements will provide a unique window to deepen our understanding of fundamental interactions and to stringently test possible extensions of our current knowledge.

The inclusive cross section for the production of a Higgs boson represents a prototypical example of experimental and theoretical synergy.
Its role in the extraction of fundamental coupling constants is key and it provides an invaluable tool to discover potential deviations from the SM. 
Experimentally it can be determined at the LHC to astounding precision. 
In order to exploit the full potential of LHC phenomenology experimental precision must be matched by equally precise theoretical prediction. 

The dominant production mechanism of a Higgs boson at the LHC is gluon fusion. 
In comparison with other processes perturbative QCD corrections to the gluon fusion cross section are large.
In order to match current and future experimental precision this simple fact demands computation of this process to very high order in perturbation theory.
Next-to-leading oder (NLO)~\cite{Dawson:1990zj,Graudenz:1992pv,Spira:1995rr} corrections to this process are available since more than two decades.
Corrections at next-to-next-to leading order (NNLO) were computed in refs~\cite{Anastasiou2002,Harlander:2002wh,Ravindran:2003um} in an effective theory (EFT) of QCD where the top quark is considered to have infinite mass and all other quarks are massless~\cite{Inami1983,Shifman1978,Spiridonov:1988md,Wilczek1977}.
In ref.~\cite{Anastasiou:2015ema} next-to-next-to-next-to leading order (N$^3$LO) corrections were computed in terms of an expansion around the production threshold of the Higgs boson.
This result marked the first computation of a Hadron collider observable to this order in perturbation theory. 
At the desired level of precision the inclusion of many sub-dominant effects, such as electro-weak corrections and quark mass effects, in a prediction for the hadron collider observable are essential.
Furthermore, a critical assessment of all sources of uncertainties is required.
A comprehensive study achieving this goal was performed in ref.~\cite{Anastasiou:2016cez} and resulted in the state of the art prediction for LHC measurements (see also refs.~\cite{deFlorian:2016spz,Harlander:2016hcx}). 

In this article we go beyond the previous approximation of the N$^3$LO corrections to the Higgs boson gluon fusion cross section in the EFT in terms of a threshold expansion and compute it exactly.
Our calculation strongly relies on various ingredients already entering the computation of ref.~\cite{Anastasiou:2015ema}. 
Specifically, we require matrix elements integrated over phase space for the production of the Higgs boson in association with up to three partons and involving up to three loops.
Purely virtual corrections were computed in ref.~\cite{Baikov:2009bg,Gehrmann2010}. Contributions with one parton in the final state and two loops were calculated in refs.~\cite{Anastasiou:2013mca,Duhr:2014nda,Dulat:2014mda,Gehrmann:2011aa,Kilgore:2013gba}. 
Matrix elements involving two final state partons and one loop (RRV) or tree level matrix elements with three final state partons (RRR) were computed for the purposes of refs.~\cite{Anastasiou:2015ema,Anastasiou:2014lda,Anastasiou:2014vaa,Li:2014afw} in terms of a threshold expansion.
Furthermore, our result relies on infrared subtraction terms formed out of convolutions~\cite{Buehler2013a,H??schele2013} of splitting functions~\cite{Moch2004,Vogt2004} and an ultraviolet counter term based on lower loop amplitudes~\cite{Anastasiou:2012kq}. Both were already computed for the purpose of ref.~\cite{Anastasiou:2015ema}.

In order to obtain our result we compute N$^3$LO corrections to the partonic cross section due to RRV and RRR matrix elements. 
The integration over the loop and final state momenta involves complicated, high-dimensional integrals.
In order to facilitate our computation we employ the framework of reverse unitarity~\cite{Anastasiou2002,Anastasiou2003,Anastasiou:2002qz,Anastasiou:2003yy,Anastasiou2004a} that allows to relate phase space integrals to cuts of loop integrals. Subsequently, we employ powerful loop integration techniques to actually compute our phase space integrals. 
In particular, we make use of integration-by-part identities~\cite{Tkachov1981,Chetyrkin1981} in order to express our integrated matrix elements in terms of a limited set of master integrals.
We then proceed to compute these master integrals using the framework of differential equations~\cite{Kotikov1991,Gehrmann2000,Henn2013}.
The solution of differential equations requires the calculation of one boundary condition per master integral. 
To obtain these boundary conditions we perform an expansion of every master integral in terms of a threshold expansion. 
We then match the individual terms in the expansion to so-called soft master integrals that were explicitly computed in refs.~\cite{Anastasiou:2013srw,Anastasiou:2015yha}.

When solving differential equations for RRR master integrals we encounter an obstruction in the form of $2\times 2$ systems of differential equations that cannot be solved by conventional means.
The solution to these systems is given in terms of elliptic integrals. 
The appearance of elliptic integrals in the computation of Feynman integrals is well established~\cite{Broadhurst:1987ei,Bauberger:1994by,Caffo:1998du,Bonciani:2016qxi,vonManteuffel:2017hms,CaronHuot:2012ab,Bourjaily:2017aa,Chen:2017soz} but still poses a considerable challenge. 
The majority of known analytic results for Feynman integrals can be expressed in terms of iterated integrals referred to as generalised poly logarithms~\cite{Goncharov1998}.
A profound understanding of their analytic properties~\cite{Goncharov:2010jf,Duhr:2011zq,Goncharov1998,Panzer:2014caa,Duhr:2012fh} has been key to the success of higher order perturbation theory.
The quest for a similar understanding of iterated integrals involving elliptic functions is subject of ongoing research and has already produced vast literature~\cite{Bloch:2013tra,Laporta:2004rb,Adams:2014vja,Adams:2015gva,Remiddi:2017har,Remiddi:2016gno,Adams:2016xah,Passarino:2016zcd,Bloch:2016izu,Adams:2017ejb,Ablinger:2017bjx,Brown:2011aa,Broedel:2015aa,Hidding:2017jkk,Bourjaily:2017bsb,Broedel:2017kkb,Broedel:2017siw}.
In particular, methods to find solutions for differential equations, the understanding of functional relations among such integrals and their analytic continuation from one kinematic regime to another are of importance.
In this article we present our pragmatic solution to the problem at hand and produce for the first time a hadron collider cross section that involves the analytic treatment of an elliptic integral.

Having obtained analytic results for all required matrix elements with different parton multiplicity in the final states we combine them to form the exact correction to the partonic Higgs boson production cross section at N$^3$LO.
We then convolute our newly obtained result and all required lower order cross section with parton distribution function in order to derive physical predictions for hadron collider cross sections.
We study in detail the deviations of our results from the previous approximation of the N$^3$LO cross section~\cite{Anastasiou:2015ema,Anastasiou:2016cez}. 
Our computation allows us to remove one source of uncertainty due to the truncation of the threshold expansion from the state of the art prediction for the Higgs boson production cross section~\cite{Anastasiou:2016cez} and we update the previous result.

This article is structured as follows: 
In section~\ref{sec:setup} we setup the notation for our computation of the inclusive Higgs boson production cross section.
Next, we discuss in detail the analytic computation of the missing RRV and RRR coefficient functions in section~\ref{seq:calc}. 
We outline the general computational framework in section~\ref{sec:gencalc}. 
We discuss the treatment of elliptic integrals found when solving differential equations in section~\ref{sec:elliptic}.
In section~\ref{sec:IIs} we introduce a class of iterated integrals that serve as the main building blocks for our final result.
Next, we describe the structure of our analytic results in section~\ref{sec:analyticres}. 
In section~\ref{sec:res} we present numerical results for the EFT Higgs boson cross section through N$^3$LO in QCD perturbation theory.
We compare our new results to previous predictions obtained with a threshold expansion in section~\ref{sec:THcompare}.
Finally, we draw our conclusions in section~\ref{sec:conclusions}.

\section{Set-Up}
\label{sec:setup}
In this article we consider scattering processes of two protons that produce at least a Higgs boson.
\begin{equation}
{\rm Proton}(P_1) + {\rm Proton}(P_2) \to H(p_h) + X,
\end{equation}
$P_1$ and $P_2$ are the momenta of the colliding protons and $p_h$ the momentum of the Higgs boson.
The master formula for the inclusive Higgs boson production cross section is given by
\beq
\label{eq:xsdiffhad2}
\sigma_{PP\rightarrow H+X}=
\tau \sum_{i,j} \int_\tau^1 \frac{dz}{z}\int_{\frac{\tau}{z}}^1 \frac{dx_1}{x_1} f_i(x_1)f_j\left(\frac{\tau}{x_1 z}\right)
\frac{1}{z}\hat\sigma_{ij}(z,m_h^2).
\eeq
Here, we employed the parton model and factorization of long and short range interactions into parton distribution functions $f_i(x)$ and partonic cross sections. 
The momenta of the colliding partons are related to the proton momenta by $p_1=x_1 P_1$ and $p_2=x_2 P_2=\frac{\tau}{x_1 z}P_2$. 
We define
\bea
\tau&=&\frac{m_h^2}{S},\hspace{1cm} S=(P_1+P_2)^2.\nonumber\\
z&=&\frac{m_h^2}{s},\hspace{1cm} s=(p_1+p_2)^2.
\eea
The sum over $i$ and $j$ ranges over all contributing partons. Furthermore, we define the variable $\bar z= 1-z$.
The partonic Higgs cross section is given by $\hat \sigma_{ij}(z,m_h^2)$. 

In this article we compute the partonic cross section through N$^3$LO in perturbative QCD in an effective theory where the top quark is infinitely heavy and has been integrated out~\cite{Inami1983,Shifman1978,Spiridonov:1988md,Wilczek1977}. 
In this theory the Higgs boson is coupled directly to gluons via an effective operator of dimension five~\cite{Chetyrkin:1997un,Schroder:2005hy,Chetyrkin:2005ia,Kramer:1996iq},
\beq
\mathcal{L}_{\text{eff}}=\mathcal{L}_{SM,5}-\frac{1}{4} C^0 H G_{\mu\nu}^a G_a^{\mu\nu}.
\eeq
where H is the Higgs field, $ G_{\mu\nu}^a $ is the gluon field strength tensor and $\mathcal{L}_{SM,5}$ denotes the SM Lagrangian with $n_f$ = 5 massless quark flavours. 
The Wilson coefficient $C^0$ is obtained by matching the effective theory to the full SM in the limit where the top quark is infinitely heavy. 

Within the effective theory, we can write the partonic cross section as
\bea 
\frac{1}{z}\hat{\sigma}_{ij}(z,\m_h^2)&=& (C^0)^2 \,\hat{\sigma}_0 \, \eta_{ij}(z)= (C^0)^2 \,\hat{\sigma}_0 \, \sum\limits_{n=0}^\infty \left(\frac{\alpha_S^0}{\pi}\right)^n \eta^{(n)}_{ij}(z).
\eea
Dividing out the Born cross section,
\beq
\hat{\sigma}_0=\frac{\pi}{8(n_c^2-1)},
\eeq
we can write the bare partonic coefficient functions as,
\bea
\tilde \eta^{(n)}_{ij}(z)&=&\frac{N_{ij}}{2 m_h^2 \hat\sigma_0}\sum\limits_{m=0}^n \int d\Phi_{H+m} \mathcal{M}^{(n)}_{ij\rightarrow H+m}.
\eea
The initial state dependent prefactors $N_{ij}$ are given by
\begin{align}
N_{gg}&=\frac{1}{4(1-\epsilon)^2(n_c^2-1)^2},\nonumber\\
N_{gq}&=N_{qg}=\frac{1}{4(1-\epsilon)(n_c^2-1)n_c},\\
N_{q\bar q}&=N_{qq}=N_{qQ}=\frac{1}{4n_c^2}.\nonumber
\end{align}
Here, $g$, $q$, $\bar q$ and $Q$ indicate that the initial state parton is a gluon, quark, anti-quark or quark of different flavour than $q$ respectively.
$d\Phi_{H+m} $ is the phase space measure for the production of a Higgs boson and $m$ partons and is explained in more detail below.
 $ \mathcal{M}^{(n)}_{ij\rightarrow H+m}$ is the coefficient of $\alpha_S^n$ in the coupling constant expansion
of the modulus squared of all amplitudes for partons $i$ and $j$ producing a final state Higgs boson and $m$ partons summed over polarizations and colors. To compute the n$^{\text{th}}$ order partonic coefficient functions we require all combinations $l$-loop matrix elements with $m$ external particles such that $m+l=n$. 

The occurring loop amplitudes are plagued by ultraviolet divergencies which we regulate using dimensional regularisation and work in $d=4-2\epsilon$ space-time dimensions.
We renormalise the Wilson coefficient and strong coupling constant in the $\overline{\text{MS}}$ scheme. 
Squared matrix elements with fixed parton multiplicity in the final state are separately infrared divergent. 
These infrared divergences are canceled by summing over all contributing squared matrix elements and performing a suitable redefinition of the parton distribution functions.
The resulting partonic cross section is free of divergencies and we refer to the corresponding partonic coefficient function as $ \eta_{ij}(z)$. 
Various definitions regarding renormalisation and mass factorisation can be found in appendix~\ref{sec:consts}.
The cross section, eq.~\eqref{eq:xsdiffhad2}, can be written in terms of finite partonic coefficient functions and physical parton distribution functions $f_i^R$ as 
\beq
\sigma_{PP\rightarrow H+X}=
\tau C^2 \hat \sigma^0 \sum_{i,j} \int_\tau^1 \frac{dz}{z}\int_{\frac{\tau}{z}}^1 \frac{dx_1}{x_1} f^R_i(x_1)f^R_j\left(\frac{\tau}{x_1 z}\right)
\eta_{ij}(z).
\eeq

The partonic coefficient functions can be split into two contributions 
\beq
\eta_{ij}^{(n)}(z)=\eta_{ij}^{(n),\text{ SV}}(z)+\eta_{ij}^{(n),\text{ reg.}}(z).
\eeq
The term $\eta_{ij}^{(n),\text{ SV}}(z)$ is comprised of distributions that act on parton distribution functions. 
The super-script SV signifies that this term represents so-called soft-virtual contributions that arise from kinematic configurations where any parton produced in conjunction with Higgs boson is soft.
The coefficient $\eta_{ij}^{(3),\text{ SV}}(z)$ was computed in ref.~\cite{Anastasiou:2014vaa} and confirmed by ref.~\cite{Li:2014afw}. 
The coefficient functions $\eta_{ij}^{(3),\text{ reg.}}(z)$ represent the so-called regular contributions. Their functional form was approximated with a power series in $1-z$ in refs.~\cite{Anastasiou:2014lda,Anastasiou:2015ema,Anastasiou2004a}. 
The main result of this article is the complete computation of the coefficient functions $\eta_{ij}^{(3),\text{ reg.}}(z)$.
We supply this result in a machine readable format in an ancillary file together with the arXiv submission of this article.

\section{Calculation of Coefficient Functions}
\label{seq:calc}

In order to obtain the partonic coefficient functions $\eta_{ij}^{(3)}(z)$ we require contributions arising from matrix elements with up to three loops ($l \leq 3$) and up to three partons ($m\leq 3$) in the final state such that $3=l+m$.
The purely virtual matrix elements were computed in refs.~\cite{Gehrmann2010,Baikov:2009bg}. 
Matrix elements with two loops and one emission were computed in refs.~\cite{Anastasiou:2013mca,Duhr:2014nda,Dulat:2014mda,Gehrmann:2011aa,Kilgore:2013gba}.
Matrix elements with two real emissions and one loop (RRV) and three real emissions (RRR) are so-far publicly only available in terms of the first two expansion terms in the expansion around the production threshold of the Higgs bosons~\cite{Anastasiou:2013srw,Anastasiou:2015yha}.
In this article we complete the computation of the N$^3$LO coefficient functions.
We start by outlining the strategy involved in this computation. 
Next, we explain the treatment of an ellitpic integral that is part of the RRR coefficient functions. 
We introduce a class of iterated integrals that serve as building blocks of our partonic coefficient functions. 
Finally, we obtain the N$^3$LO coefficient functions and describe their structure.

\subsection{Computation of Matrix Elements}
\label{sec:gencalc}
In order to obtain RRV and RRR coefficient functions we start by generating all required Feynman diagrams with QGRAF~\cite{Nogueira1993}.
Next, we perform spinor and colour algebra in a private c++ code based on GiNaC~\cite{Bauer2000}.
 With this we obtain the loop and phase space integrand for our partonic coefficient functions.
 
 Next, we want to perform the inclusive integral of our integrands over all loop momenta and final state parton momenta.
 The phase space measure for producing a Higgs boson and $m$ partons is given by
 \beq
 \label{eq:measure}
d\Phi_{H+m}= \frac{d^dp_h}{(2\pi)^d} (2\pi)\delta_+(p_h^2-m_h^2)(2\pi)^d \delta^d\left(p_1+p_2+p_h+\sum\limits_{i=3}^{m+2} p_i\right)\prod\limits_{i=3}^{m+2}\frac{d^dp_i}{(2\pi)^d} (2\pi)\delta_+(p_i^2),
 \eeq
 where 
\beq
\delta_+(p^2-m^2)=\theta(-p^0-m)\delta(p^2-m^2).
\eeq
All final state momenta are chosen in-going such that the energy component in the above equation appears with a minus sign.
In order to perform the loop and phase space integration we 
employ the framework of reverse unitarity~\cite{Anastasiou2002,Anastasiou2003,Anastasiou:2002qz,Anastasiou:2003yy,Anastasiou2004a} 
that allows to treat phase space and loop integrals on equal footing.
In particular, we represent the on-shell constraints in terms of cut propagators.
\beq
\delta_+(p^2-m^2)\rightarrow \left[\frac{1}{p^2-m^2}\right]_c
\eeq
The subscript c serves as a reminder that this propagator is cut. Cut propagators can be differentiated just like normal propagators.
\beq
\frac{d}{dx}\left[\frac{1}{f(x)}\right]_c^{a}=-a \left[\frac{1}{f(x)}\right]_c^{a+1}\frac{df(x)}{dx}.
\eeq
They satisfy the condition 
\beq
 \left[\frac{1}{f(x)}\right]_c^a f(x)^b=\left\{ \begin{array}{cc}  \left[\frac{1}{f(x)}\right]_c^{a-b}&,\text{if }a>b \\ 0 &,\text{if }b\geq a\end{array}\right. .
\eeq
We can now apply integration-by-part (IBP) identities~\cite{Tkachov1981,Chetyrkin1981} on our combined loop and phase-space integrands.
A private c++ implementation of the Laporta algorithm~\cite{Laporta:2001dd} allows us to express our partonic coefficient functions in terms of a limited set of master integrals.
To compute these master integrals we work with the method of differential equations~\cite{Kotikov1991,Gehrmann2000,Henn2013}. 
This method allows to derive a system of partial differential equations for a vector of our master integrals $\vec{I}(z)$ of the form
\beq
\frac{\partial}{\partial z} \vec{I}(z) =A(z,\epsilon) \vec{I}(z).
\eeq
Here, $\vec{I}(z)$ is a vector of $n$ master integrals and $A(z,\epsilon)$ is a $n\times n$ matrix with ratios of polynomials in $z$ and $\epsilon$ as entries. 
In order to have a complete system of differential equations we define 550 and 362 master integrals for RRR and RRV respectively.
 
 The commonly used strategy to solve such differential equations is to find a $n\times n$ transformation matrix $T$ such that 
 \bea
 \label{eq:canform}
 \vec{I}(z)&=&T \vec{I}^\prime(z).\nonumber\\
\epsilon A^\prime(z,\epsilon) &=&T^{-1} A(z,\epsilon)T-T^{-1}\frac{\partial}{\partial z} T .\nonumber\\
\frac{\partial}{\partial z} \vec{I}^\prime(z) &=&\epsilon A^\prime(z,\epsilon) \vec{I}^\prime (z).
 \eea
Here, $A^\prime(z,\epsilon)$ is holomorphic in $\epsilon$ as $\epsilon\rightarrow 0$.
Having obtained such a form the solution for our master integrals can be easily expressed in terms of a Laurent series in $\epsilon$ by
\beq
\label{eq:diffsol}
\vec{I}^\prime(z)=\left[\mathbb{I}+\epsilon \int^z dz^\prime A^\prime(z^{\prime},\epsilon)+\epsilon^2 \int^z dz^\prime \int^{z^{\prime}} dz^{\prime \prime} A^\prime(z^{\prime},\epsilon)A^\prime(z^{\prime \prime},\epsilon)+\dots \right] \vec{I}^{\prime}_0.
\eeq
Here, $\vec{I}^{\prime}_0$ represents a vector of boundary conditions that has to be determined by other means. 
For the RRV and RRR master integrals such a boundary condition is easily obtained by matching the full solution obtained in eq.~\ref{eq:diffsol} to an expansion of the required integrals $\vec{I}(z)$ around the point $z=1$ that can be performed by means presented in ref.~\cite{Anastasiou:2013srw,Anastasiou:2015yha}.

The art in solving differential equations rests in finding an adequate transformation matrix $T$. 
For certain differential equations in a single parameter an algorithmic solution exists~\cite{Barkatou:2007:CSF:1277548.1277550,BARKATOU20091017,Moser1959,Lee:2014ioa} and was nicely formulated in ref.~\cite{Lee:2014ioa}.
This method applies if a transformation matrix can be found that is comprised of ratios of polynomials in the parameters $z$ and $\epsilon$. 
For a large subset of integrals in our vector of master integrals $\vec{I}$ such transformations can be found and we rely on a private implementation of the algorithm outlined in ref.~\cite{Lee:2014ioa} to do so.

For another large class of master integrals it is necessary to find a transformation matrix that contains square roots of polynomials of our parameter z.
For these cases we can find the desired transformation by finding suitable algebraic variable transformations that rationalises the square roots involved. 
Once the roots are rationalised we can again employ the aforementioned algorithm.
We point out that this procedure is not particularly algorithmic but leads to a desired solution fairly easily.

We encounter a further obstruction when solving differential equations for the system of RRR master integrals. 
This obstruction involves the presence of elliptic integrals and we elaborate on our solution in the following section.

\subsection{An Elliptic Integral in Higgs Production}
\label{sec:elliptic}
\begin{figure*}[!ht]
\centering
\begin{subfigure}[b]{0.47\textwidth}
\includegraphics[width=0.95\textwidth]{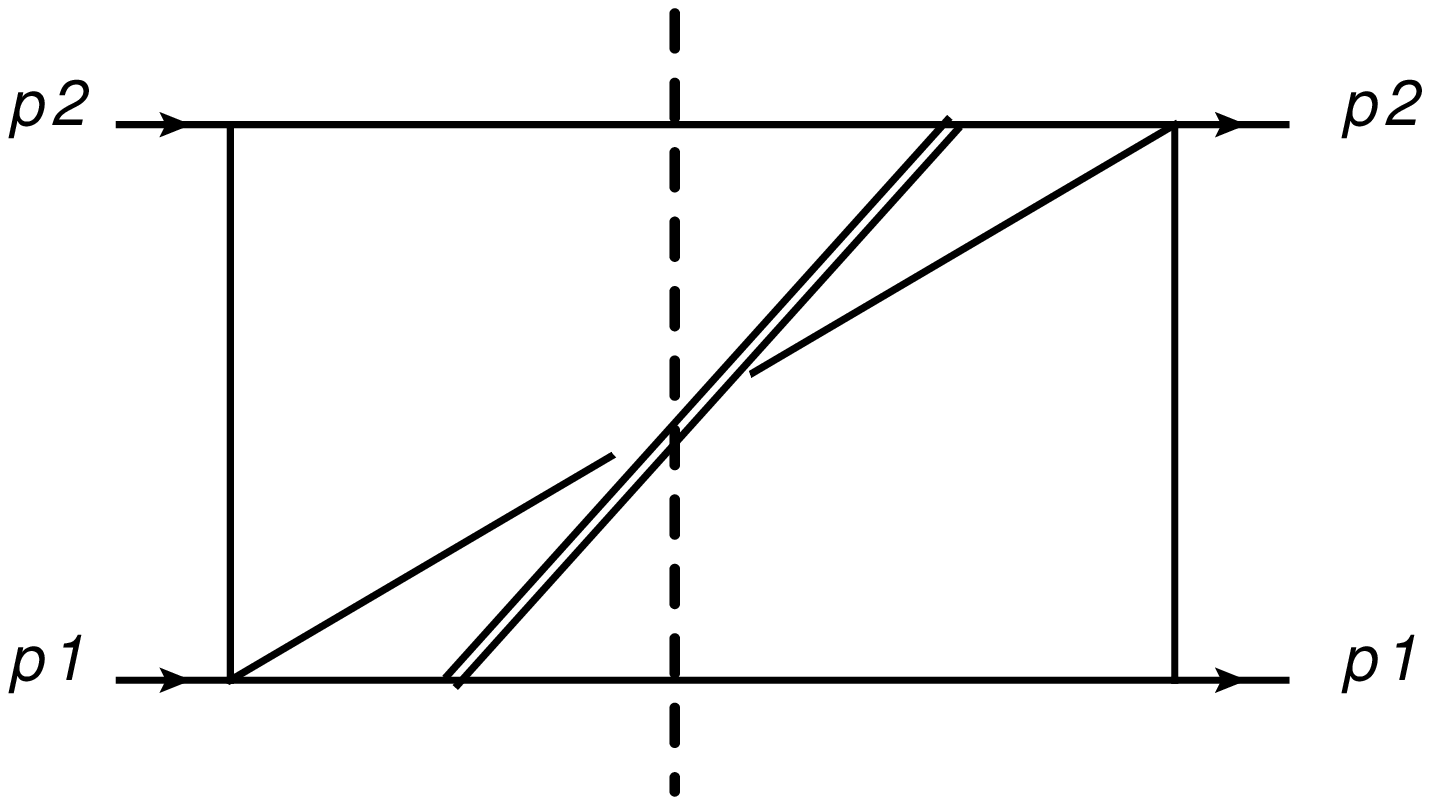}
\caption{ \label{fig:elliptica}
}
\end{subfigure}
\begin{subfigure}[b]{0.47\textwidth}
\includegraphics[width=0.95\textwidth]{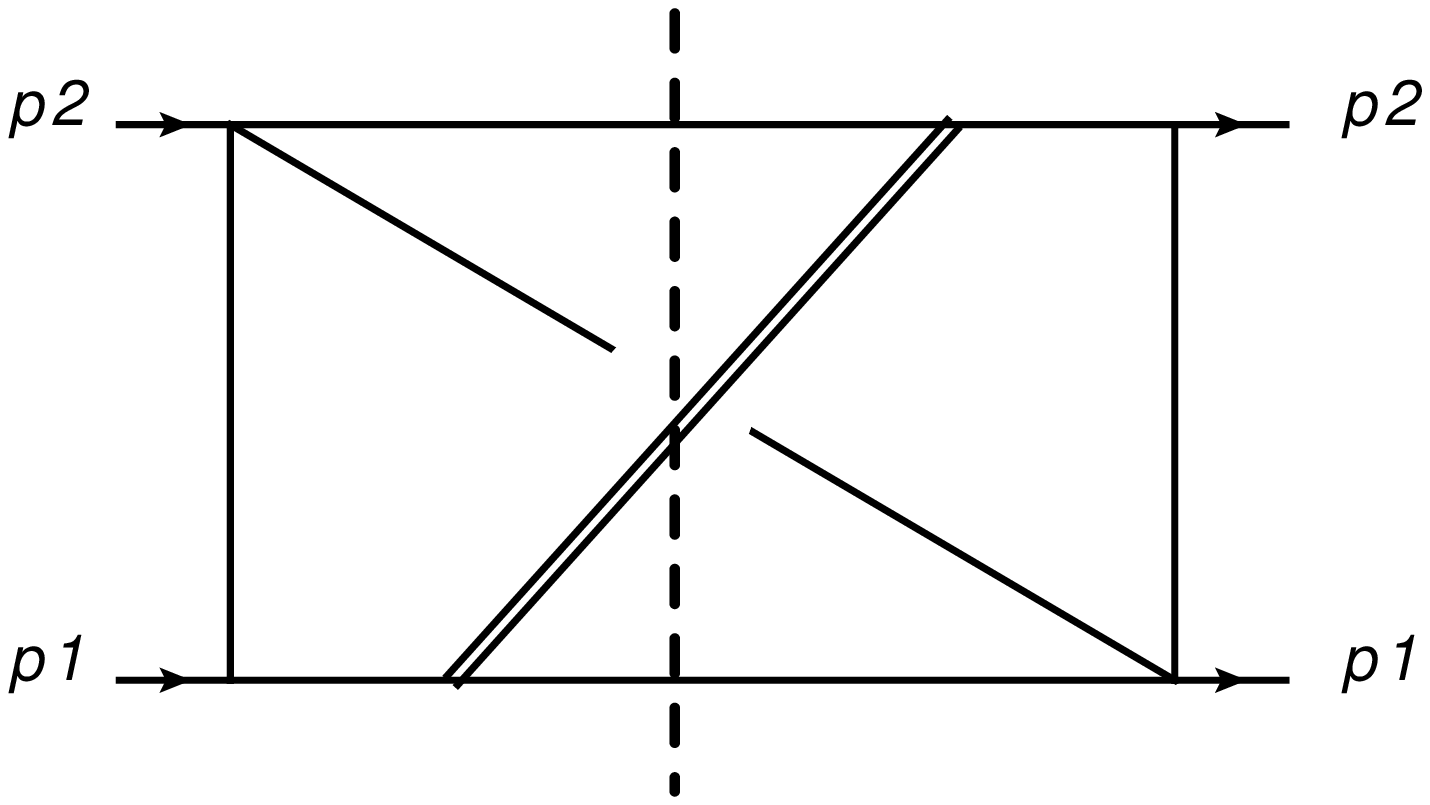}
\caption{ \label{fig:ellipticb}
}
\end{subfigure}
\caption{\label{fig:elliptic}
Phase space integrals contributing to triple real corrections to Higgs boson production at N$^3$LO. 
The computation of these integrals involves elliptic integrals. 
Solid lines represent Feynman propagators. Solid lines crossed by the dashed line correspond to cut-propagators. 
The doubled line represents the on-shell constraint of the Higgs boson.
}
\end{figure*}

When solving differential equations for master integrals contributing to the triple real coefficient functions of Higgs boson production at N$^3$LO we encounter two coupled $4\times 4$ systems of differential equations that we could not decouple order by order in the dimensional regulator by conventional means. 
In this section we discuss the differential equations in question and present our solution.

In figure~\ref{fig:elliptic} we display two scalar phase space integrals. 
Let us choose four master integrals with the same propagators as the scalar integral in figure~\ref{fig:ellipticb}.
\beq
E_i=\int d\Phi_{H+3}\frac{n_i}{p_{145}^2p_{235}^2p_{1245}^2 p_{1235}^2},\hspace{1cm} p_{i_1\dots i_n}=p_{i_1}+\dots+p_{i_n}.
\eeq
We choose
\bea
n_1&=&\frac{z s^3}{\epsilon (p_{12345}^2-s z)}.\nonumber\\
n_2&=&-\frac{s}{16}(p_{14}^2+p_{23}^2+p_{35}^2).\nonumber\\
n_3&=&-\frac{s}{16}(p_{23}^2+p_{35}^2).\nonumber\\
n_4&=&\frac{s^2}{\epsilon}.\nonumber\\
\eea
These four integrals satisfy a system of differential equations of the form
\beq
\label{eq:elldiff}
\frac{\partial}{\partial z}\vec{E} = A_0(z) \vec{E}+\epsilon A_1(z,\epsilon) \vec{E}+\vec{y}(z).
\eeq
The vector $\vec{y}(z)$ represents the inhomogeneous part of the differential equation. 
The matrix $A_1(z,\epsilon)$ in the homogeneous part of the differential equation is holomorphic in $\epsilon$ as $\epsilon \rightarrow 0$.
The homogeneous part of the differential equation that does not decouple as $\epsilon \rightarrow 0$ is given by the matrix
\beq
\label{eq:A0Def}
A_0(z)=\left(
\begin{array}{cccc}
 \frac{11-2 z}{z^2-11 z-1} & 0 & 0 & \frac{3-z}{z^2-11 z-1} \\
 0 & 0 & 0 & 0 \\
 0 & 0 & 0 & 0 \\
\frac{1}{z} & 0 & 0 & 0 \\
\end{array}
\right).
\eeq
As we can see, for $\epsilon=0$ two of the master integrals decouple and we are left with a $2 \times 2$ system for the homogeneous solution of the differential equation. 
\beq
\label{eq:elldiff2}
\frac{\partial}{\partial z} \left(\begin{array}{c} E_4^0 \\ E_1^0 \end{array}\right)=
A_T. \left(\begin{array}{c} E_4^0 \\ E_1^0 \end{array}\right)=
\left(
\begin{array}{cc}
 0 &\frac{1}{z}   \\
\frac{3-z}{z^2-11 z-1}  &  \frac{11-2 z}{z^2-11 z-1}\\
\end{array}
\right).
 \left(\begin{array}{c} E_4^0 \\ E_1^0 \end{array}\right).
\eeq
In order to decouple our original system of eq.~\eqref{eq:elldiff} we want to find a transformation matrix $T_E$ such that
\bea
 \left(\begin{array}{c} E_4^0 \\ E_1^0 \end{array}\right)&=&T_E. \left(\begin{array}{c} E_4^{\prime 0} \\ E_1^{\prime 0} \end{array}\right)
=\left(\begin{array}{cc} t_{11}(z) & t_{12}(z) \\ t_{21}(z) & t_{22}(z) \end{array}\right) . \left(\begin{array}{c} E_4^{\prime 0} \\ E_1^{\prime 0} \end{array}\right).\nonumber\\
\frac{\partial}{\partial z}\left(\begin{array}{c} E_4^{\prime 0} \\ E_1^{\prime 0} \end{array}\right) &=&0.\nonumber\\
\frac{\partial}{\partial z} T_E&=&A_T. T_E.
\label{eq:tdef}
\eea
We show in appendix~\ref{sec:APPEll} that the functions $t_{ij}(z)$ can be written in terms of complete elliptic integrals and pre-factors.  
However, this solution is quite unwieldy and we choose another approach here.
For all practical purposes it is sufficient to simply define the functions $t_{ij}(z)$ 
to be the solution to the differential equation eq.~\eqref{eq:tdef}. 
The homogeneous differential equations for the master integrals $E_1^\prime$ and $E_4^\prime$, defined by
\bea
E_1&=&t_{22} E_1^\prime +t_{21} E_4^\prime,\nonumber\\
E_4&=&t_{11} E_4^\prime +t_{12} E_1^\prime,
\eea
are decoupled as we send $\epsilon\rightarrow 0$. 
The inhomogeneity can then be decoupled order by order by in $\epsilon$ by standard techniques.
A general solution for the differential equations can subsequently be found as illustrated by eq.~\eqref{eq:diffsol}. 

The second set of master integrals that have the same propagators as the scalar integral depicted in figure~\ref{fig:elliptica} can be chosen in such a way that the homogeneous part of their differential equations takes identically the same form as the one already discussed. 
Therefor we can apply the same transformation matrix to decouple the system order by order in $\epsilon$.
With this we found a transformation matrix $T$ that allows us to express the differential equations for all master integrals required for RRV and RRR contributions to Higgs production at N$^3$LO in the desired form, eq.~\eqref{eq:canform}.

In order to derive numerical results for the functions $t_{ij}$ we can solve the differential equations eq.~\eqref{eq:tdef} in terms of a generalised power series ansatz using the Frobenius method. 
Consider for example an ansatz for the solution of the system of differential equations as an expansion around $z=0$ and $z=1$.
\bea
t_{ij}(z)&=&\sum_{n=0}^\infty \bar z^n b_{ij}^{(n)}.\nonumber\\
t_{ij}(z)&=&\sum_{n=0}^\infty  z^n c_{ij}^{(n)}+\log(z)\sum_{n=0}^\infty d_{ij}^{(n)}z^n.
\eea
We derived the required structure of our ansatz by regarding the asymptotic solution of the differential equations around the considered expansion points.
\bea
\label{eq:asysol}
T_E&=&e^{-\log(\bar z)  \lim\limits_{z\rightarrow 1} \bar z A_T }.\left(\begin{array}{cc}t^{1}_{11} & t^{1}_{12}\\t^{1}_{21} & t^{1}_{22} \end{array}\right) =\left(\begin{array}{cc}t^{1}_{11} & t^{1}_{12}\\t^{1}_{21} & t^{1}_{22} \end{array}\right)+\mathcal{O}(\bar z^1).\nonumber\\
T_E&=&e^{\log( z)  \lim\limits_{z\rightarrow 0} z A_T }.\left(\begin{array}{cc}t^{0}_{11} & t^{0}_{12}\\t^{0}_{21} & t^{0}_{22} \end{array}\right)=
\left(\begin{array}{cc}t^{0}_{11} & t^{0}_{12}\\t^{0}_{21} & t^{0}_{22} \end{array}\right)
+\log(z)\left(\begin{array}{cc}t^{0}_{21} & t^{0}_{22} \\ 0 & 0 \end{array}\right)+\mathcal{O}(z^1).
\eea
Here, $t^0_{ij}$ and $t^1_{ij}$ are some numerical boundary constants.

Inserting the ansaetze into the system of differential equations we find the following recurrence relations.
\bea
\label{eq:rec1}
b_{11}^{(n+2)}&=&\frac{(n+1) b_{11}^{(n+1)}}{n+2}-\frac{b_{21}^{(n+1)}}{n+2}.\nonumber\\
b_{21}^{(n+2)}&=&\frac{b_{11}^{(n)}}{11 (n+2)}+\frac{2 b_{11}^{(n+1)}}{11 (n+2)}+\frac{1}{11} b_{21}^{(n)}+\frac{9}{11} b_{21}^{(n+1)}.
\eea
and
\bea
\label{eq:rec2}
c_{11}^{(n+2)}&=& \frac{c_{11}^{(n)}}{(n+2)^2}-\frac{3 c_{11}^{(n+1)}}{(n+2)^2}+\frac{c_{21}^{(n)}}{n+2}-\frac{11 c_{21}^{(n+1)}}{n+2}-\frac{2 d_{11}^{(n)}}{(n+2)^3}\nonumber\\
&+&\frac{6 d_{11}^{(n+1)}}{(n+2)^3}-\frac{d_{21}^{(n)}}{(n+2)^2}+\frac{11 d_{21}^{(n+1)}}{(n+2)^2}.\nonumber\\
c_{21}^{(n+2)}&= &\frac{c_{11}^{(n)}}{n+2}-\frac{3 c_{11}^{(n+1)}}{n+2}+c_{21}^{(n)}-11 c_{21}^{(n+1)}-\frac{d_{11}^{(n)}}{(n+2)^2}+\frac{3 d_{11}^{(n+1)}}{(n+2)^2}.\nonumber\\
d_{11}^{(n+2)}&=& \frac{d_{11}^{(n)}}{(n+2)^2}-\frac{3 d_{11}^{(n+1)}}{(n+2)^2}+\frac{d_{21}^{(n)}}{n+2}-\frac{11 d_{21}^{(n+1)}}{n+2}.\nonumber\\
d_{21}^{(n+2)}&=& \frac{d_{11}^{(n)}}{n+2}-\frac{3 d_{11}^{(n+1)}}{n+2}+d_{21}^{(n)}-11 d_{21}^{(n+1)}.
\eea
By comparison to the asymptotic solution given in eq.~\eqref{eq:asysol} we find all starting conditions for the solution to the recurrence relations.
Specifically, we find the conditions  $b_{ij}^{(n)}=c_{ij}^{(n)}=d_{ij}^{(n)}=0$ if $n<0$ and $d_{21}^{(0)}=0$ and $d_{11}^{(0)}=c_{21}^{(0)}$. 
Furthermore, the general solution for $t_{22}$ is identical to the solution for $t_{21}$ and the one for $t_{12}$ is identical to the solution for $t_{11}$ up to the choice of boundary constants. 

Any choice of boundary conditions will lead to a transformation matrix that satisfies the differential equations eq.~\eqref{eq:tdef}.
The only restriction we impose is that the transformation has to be invertible, i.e. that $\det(T_E)\neq 0$. In accordance with this criterium we make the following choice for the asymptotic solution of the differential equation: $t^1_{11}=t^1_{22}=0$ and $t^1_{12}=t^1_{21}=1$. We find that with this choice the determinant of the transformation matrix is given to all orders in $z$ by
\beq
t_{11} t_{22}-t_{12}t_{21}=-\frac{11}{z^2-11 z-1}.
\eeq

Fixing the asymptotic behaviour of the functions $t_{ij}(z)$ in one limit automatically determines their behaviour at any other point.
Computing the explicit values for $t_{ij}^0$ explicitly given the choice we made for $t_{ij}^1$ is however a non-trivial task. 
%Such a solution would require us to find a some closed analytic form for the solution to the differential equations. 
At this point it is useful to reflect on the practical aim of our computation. 
We desire a solution that is numerically sufficiently precise to determine the complete N$^3$LO coefficient functions for values of $z$ in the interval $[0,1]$ as required for cross section predictions. 
In this light our solution for the $t_{ij}(z)$ should allow for the desired precision and should be improvable if necessary.
This can be achieved by computing an approximation based on a truncated power series.

The regular singular points of our $2\times 2$ system of differential equations ~\eqref{eq:tdef} are located at the following values.
\bea
z&=&0.\nonumber\\
z&=&\frac{1}{2} \left(11-5 \sqrt{5}\right)\sim-0.09.\nonumber\\
z&=&\frac{1}{2} \left(11+5 \sqrt{5}\right)\sim11.09.
\eea
Consequently, the power series of the functions $t_{ij}(z)$ around the point $z=1$ has a radius of convergence $r_1=1$. Similarly, the power series around the point $z=0$ has radius of convergence $r_0=|\frac{1}{2} \left(11-5 \sqrt{5}\right)|$.
The domains of convergence for the two power series overlap on the interval $z\in(0,|\frac{1}{2} \left(11-5 \sqrt{5}\right)|)$.
In order to determine the boundary constants $t^0_{ij}$ in terms of the $t^1_{ij}$ we first compute the truncated power series around both limits under consideration for each $t_{ij}(z)$. Next, we evaluate both series for each $t_{ij}(z)$ at a point within the interval $z\in(0,|\frac{1}{2} \left(11-5 \sqrt{5}\right)|)$. Equating the results allows us to establish a relation among the constants $t^0_{ij}$ and $t^1_{ij}$ up to a small, numerical remainder. The remainder can be systematically improved upon by increasing the truncation order of the power series. 

Let us briefly introduce a simple method of estimating the size of the remainder of the truncated series.
Suppose a function $f(x)$ is given by the convergent series
\beq
f(x)=\sum\limits_{i=0}^\infty a_i x^i.
\eeq
If we truncate the series before order $N$ its remainder would be given by 
\beq
R(f(x),x,N)=\sum\limits_{i=N}^\infty a_i x^i.
\eeq
Suppose that asymptotically the ratio of to consecutive coefficients remains constant.
\bea
|a_{i+1}|&=&|a_i| r_i\nonumber\\
|a_{i+m}|&=& r_i^m | a_i |.
\eea
Under this assumption we can estimate the modulus of the remainder to be bounded by
\beq
\label{eq:restdef}
|R(f(x),x,N)|\leq |a_N| x^N \sum\limits_{i=0}^\infty (r_N x)^i= \frac{a_N x^N}{1-r_N x} = \text{Rest}(f(x),x,N).
\eeq
Note, that the series converges for $|r_N x|<1$. 

In order to obtain sufficiently high precision for our coefficient functions we perform an expansion of the functions $t_{ij}$ around the expansion points $z=0$, $z=1$ and $z=\frac{1}{2}$. 
For each expansion we compute several hundred terms and match the boundary conditions within the overlaps of the domains of convergence. 
Estimating the remainder of the power series expansion at our matching points suggests that we can easily determine the boundary values with a relative accuracy of $10^{-42}$ or better if needed. 
In addition to estimating the remainder as described above we evaluate the different power series for the same $t_{ij}$ for several points in the intervals where all series converge and only observe relative deviations at levels smaller than $10^{-42}$.

In order to further study the convergence of our power series approximation we may regard the asymptotic behaviour of the recurrence relations given in eq.~\eqref{eq:rec1} and eq.~\eqref{eq:rec2} as $n\rightarrow \infty$.
\bea
b_{11}^{(n+2)}&=&b_{11}^{(n+1)}+\mathcal{O}\left(\frac{1}{n}\right).\nonumber\\
b_{21}^{(n+2)}&=&\frac{1}{11} b_{21}^{(n)}+\frac{9}{11} b_{21}^{(n+1)}+\mathcal{O}\left(\frac{1}{n}\right).\nonumber\\
c_{11}^{(n+2)}&=&0+\mathcal{O}\left(\frac{1}{n}\right).\nonumber\\
c_{21}^{(n+2)}&= &c_{21}^{(n)}-11 c_{21}^{(n+1)}+\mathcal{O}\left(\frac{1}{n}\right). \nonumber\\
d_{11}^{(n+2)}&=&0+\mathcal{O}\left(\frac{1}{n}\right).\nonumber\\
d_{21}^{(n+2)}&=& d_{21}^{(n)}-11 d_{21}^{(n+1)}+\mathcal{O}\left(\frac{1}{n}\right).\nonumber\\
\eea
We see that asymptotically $b_{11}^{(n)}$ approaches a constant and $c_{11}^{(n)}$ and $d_{11}^{(n)}$ tend towards zero. For the other coefficients we find the asymptotic solutions
\bea
b_{21}^{(n)}&=&\left(\frac{9}{22} - \frac{5 \sqrt{5}}{22}\right)^n c_1 + \left(\frac{9}{22} + \frac{5 \sqrt{5}}{22}\right)^ n c_2+\mathcal{O}\left(\frac{1}{n}\right).\nonumber\\
c_{21}^{(n)}&=&\left(-\frac{11}{2} - \frac{5 \sqrt{5}}{2}\right)^n c_3 +\left(-\frac{11}{2} + \frac{5 \sqrt{5}}{2}\right)^ n c_4+\mathcal{O}\left(\frac{1}{n}\right).\nonumber\\
d_{21}^{(n)}&=&\left(-\frac{11}{2} - \frac{5 \sqrt{5}}{2}\right)^n c_5 + \left(-\frac{11}{2} + \frac{5 \sqrt{5}}{2}\right)^ n c_6+\mathcal{O}\left(\frac{1}{n}\right).
\eea
Here, the $c_i$ are some numerical constants. The numbers $\left|\frac{9}{22} \pm \frac{5 \sqrt{5}}{22}\right|$ and $\left|-\frac{11}{2} + \frac{5 \sqrt{5}}{2}\right|$ are smaller than one.
The number $\left|-\frac{11}{2} - \frac{5 \sqrt{5}}{2}\right|$ is larger than one. From this we again draw the conclusion that the power series around the expansion point $z=1$ is convergent everywhere within the unit interval.
The power series around $z=0$ is convergent if $z<1/\left|-\frac{11}{2} - \frac{5 \sqrt{5}}{2}\right|=\left| \frac{1}{2} \left(11-5 \sqrt{5}\right)\right|$.
This asymptotic analysis also supports the validity of the procedure to estimate the remainder of the power series truncated at order $N$ defined in eq.~\eqref{eq:restdef}.

\subsection{Iterated Integrals}
\label{sec:IIs}
In this section we briefly introduce a class of iterated integrals~\cite{Chen:1977oja} that is particularly convenient to express the solution of differential equations as in  eq.~\eqref{eq:diffsol}.
We define 
\beq
\label{eq:iidef}
J(\vec{\omega},z)=J(\omega_n(z),\dots,\omega_1(z), z)=\int^z_0 dz^\prime \omega_n(z^\prime) J(\omega_{n-1}(z^\prime),\dots,\omega_1(z^\prime),z^\prime),
\eeq
with $J(z)=1.$
We refer to one $\omega_i(z)$ as a letter and to an ordered set of letters, $\{\omega_n(z),\dots,\omega_1(z)\}$ that defines an iterated integral as a word. 

Many well known classes of iterated integrals, such as harmonic poly logarithms (HPLs) ~\cite{REMIDDI2000} or generalised poly logarithms (GPLs)~\cite{Goncharov1998}, that are widely used in particle physics, are sub-classes of this type of iterated integrals. 
For example the GPLs are given by
\beq
G(a_n,\dots,a_1,z)=J\left(\frac{1}{z-a_n},\dots,\frac{1}{z-a_1},z\right),\hspace{1cm}a_i\in\mathbb{C}.
\eeq
The presence of the elliptic integrals $t_{ij}(z)$ in the solution of our differential equations does not allow for a solution purely in terms of GPLs.
For this reason it becomes necessary to define an extension of GPLs in this article.
Already several generalisations of GPLs to accommodate elliptic functions exist in the literature (see for example~\cite{Adams:2016sob,Broedel:2016wiz,Remiddi:2017har,Ablinger:2014wca,Broedel:2017kkb,Broedel:2017siw}).
In the following we review several properties of iterated integrals (see for example~\cite{Duhr:2011zq,Panzer:2014caa,Panzer:2014gra,Duhr:2012fh}).

Iterated integrals form a so-called shuffle algebra.
  \beq\bsp\label{eq:G_shuffle}
  J(\omega_n(z),\ldots,\omega_{1}(z);z) \, J(\omega_{n+m}(z),\ldots,\omega_{n+1}(z),z) &\,=\sum_{\sigma\in\Sigma(n,m)}\,J(\omega_{\sigma(n+m)}(z),\ldots,\omega_{\sigma(1)}(z),z),\\
      \esp
\eeq
where $\Sigma(n,m)$ denotes the set of all shuffles of $n+m$ elements, \emph{i.e.}, the subset of the symmetric group $S_{n+m}$ defined by
\beq\bsp\label{eq:Sigma_def}
&\Sigma(n,m) =\{\sigma\in S_{n+m} |\, \sigma^{-1}(n)<\ldots<\sigma^{-1}({1}) {\rm~~and~~} \sigma^{-1}(n+m)<\ldots<\sigma^{-1}(n+1)\}\,.
\esp\eeq 
For example, consider the product of two iterated integrals with two integrations each.
\bea
J(a,b,z) J(c,d,z)&=&
J(a,b,c,d,z)+J(a,c,b,d,z)+J(a,c,d,b,z)\nonumber\\
&+&J(c,a,b,d,z)+J(c,a,d,b,z)+J(c,d,a,b,z).
\eea
Here the letters $a$, $b$, $c$ and $d$ may be generic functions of $z$.

Special care needs to be taken if the integrand of our iterated integrals diverges at the value of the lower integration bound.
In this article we only consider simple poles of the integrand at the end points since they simply are the only type of divergence that appears in the computation we are interested in.
Specifically, we define the case where all letters of a word of lenght $n$ are given by $\omega(z)=\frac{1}{z}$ then
\beq
J\left(\frac{1}{z},\dots,\frac{1}{z},z\right)=\frac{1}{n!} \log^n(z).
\eeq
If the letter $\frac{1}{z}$ appears in the right-most entry of the word of an iterated integral we define it in a way that is consistent with the shuffle algebra.
Consider the shuffle relation
\beq
J\left(\frac{1}{z},z\right)  J\left(\omega_n(z),\dots,\omega_1(z),z\right) =J\left(\omega_n(z),\dots,\omega_1(z),\frac{1}{z},z\right) +J\left(\omega_n(z),\dots,\frac{1}{z},\omega_1(z),z\right) +\dots.
\eeq
Here, the ellipsis indicates all other terms arising from the shuffle product. 
Assuming that all $\omega_i(z)$  in the above equation are holomorphic as $z\rightarrow 0$ the only iterated integral with an end-point divergence is the first on the right hand side of the equation.
We define our iterated integrals to be regulated in such cases such that the above equation holds true. Solving for the iterated integral in question we find
\beq
J\left(\omega_n(z),\dots,\omega_1(z),\frac{1}{z},z\right) =\log(z)J\left(\omega_n(z),\dots,\omega_1(z),z\right)  - J\left(\omega_n(z),\dots,\frac{1}{z},\omega_1(z),z\right) +\dots.
\eeq
If the right-most letter is divergent as $z\rightarrow 0$ but has the form $\frac{f(z)}{z}$, with f(z) being holomorphic around $z=0$, then we may regularise our function by writing it as
\bea
J\left(\omega_n(z),\dots,\omega_1(z),\frac{f(z)}{z},z\right)&=&J\left(\omega_n(z),\dots,\omega_1(z),\frac{f(z)-f(0)}{z},z\right)\nonumber\\
&+&J\left(\omega_n(z),\dots,\omega_1(z),\frac{1}{z},z\right)f(0).
\eea
The last line of the above equation is then regulated as discussed above.
If several right-most letters have poles at the lower end point of the integration we simply iterate the above procedure.

We want to be able to rewrite an elliptic integral with argument $z$ in terms of iterated integrals with argument $\bar z=1-z$  or $w=\frac{1}{2}-z$. 
Let us illustrate how this can be achieved by regarding a transformation from $z$ to $\bar z$.
\bea
\label{eq:rewriteJ}
J(\omega_n(z),\dots,\omega_1(z), z)&=&\int^{1-\bar z}_0 dz^\prime \omega_n(z^\prime) J(\omega_{n-1}(z^\prime),\dots,\omega_1(z^\prime),z)\nonumber\\
&=&-\int^{\bar z}_1 d\bar z^\prime \omega_n(1-\bar z^\prime) J(\omega_{n-1}(1-\bar z^\prime),\dots,\omega_1(1-\bar z^\prime),1-\bar z^\prime)\nonumber\\
&=&-\int^{\bar z}_0 d\bar z^\prime \omega_n(1-\bar z^\prime) J(\omega_{n-1}(1-\bar z^\prime),\dots,\omega_1(1-\bar z^\prime),1-\bar z^\prime)\nonumber\\
&+&\int^{1}_0 d\bar z^\prime \omega_n(1-\bar z^\prime) J(\omega_{n-1}(1-\bar z^\prime),\dots,\omega_1(1-\bar z^\prime),1-\bar z^\prime).
\eea
The last line in the above equation is a numerical constant.
In order to write the integral in the penultimate line in terms of an iterated integral with upper integration bound $\bar z$ we have to first rewrite the iterated integral in the integrand  with an upper integration bound $\bar z ^\prime$.
To do this we simply apply the above equation iteratively to the integrand.
Notice, that the above procedure may be ill defined if the integrand we are considering is divergent at any of the end points. 
This case is easily avoided by shuffle regulating both end points prior to applying eq.~\eqref{eq:rewriteJ}. 
Let us demonstrate this step with a well known example. Consider the iterated integral
\bea
\label{eq:int1}
J\left(\frac{1}{z},\frac{1}{1-z},z\right)&=&J\left(\frac{1}{ z}, z\right)J\left(\frac{1}{ 1-z}, z\right)-J\left(\frac{1}{ 1-z},\frac{1}{ z}, z\right)\nonumber\\
&=&-\log(z)\log(1-z)-J\left(\frac{1}{ 1-z},\frac{1}{ z}, z\right)
\eea
In the above equation we employed a shuffle identity such that right most letter of the function is regular at the new lower integration point $z=1$ and that the left most letter is regular at the new end point $z=0$.
We now may write 
\bea
\label{eq:int2}
J\left(\frac{1}{ 1-z},\frac{1}{ z}, z\right)&=&\int_0^z dz^\prime \frac{1}{1-z^\prime }J\left(\frac{1}{ z^\prime }, z^\prime \right)=\int_0^z dz^\prime \frac{\log(z^\prime )}{1-z^\prime }\nonumber\\
&=&\int_0^{\bar z} d\bar z^\prime \frac{1}{\bar z^\prime }J\left(\frac{1}{ 1-\bar z^\prime }, \bar z^\prime \right)-\int_0^{1} d\bar z^\prime \frac{1}{\bar z^\prime }J\left(\frac{1}{ 1-\bar z^\prime }, \bar z^\prime \right)\nonumber\\
&=&J\left( \frac{1}{\bar z },\frac{1}{ 1-\bar z }, \bar z \right)-\frac{\pi^2}{6}.
\eea
Combining the the results of eq.~\eqref{eq:int1} and eq.~\eqref{eq:int2} we find the famous di-Logarithm identity.
\beq
J\left(\frac{1}{z},\frac{1}{1-z},z\right)=-J\left( \frac{1}{\bar z },\frac{1}{ 1-\bar z }, \bar z \right)-\log(\bar z)\log(1-\bar z)+\frac{\pi^2}{6}.
\eeq
In this example it was possible to determine the integration constant to be $\frac{\pi^2}{6}$ analytically. 
If this is not possible the constant can also be determined numerically with finite precision by simply evaluating the function under consideration before and after variable transformation numerically for any value of $z$.

The iterated integral representation of eq.~\eqref{eq:iidef} allows to easily compute truncated power series expansions for the iterated integrals.
For example 
\beq
\label{eq:serex}
J\left(\frac{1}{1-\bar z},\bar z\right)=\int^{\bar z}_0 \frac{d\bar z^\prime}{1-\bar z}=\int^{\bar z}_0 d\bar z^\prime \sum\limits_{i=0}^\infty (\bar z^\prime)^i=\sum\limits_{i=0}^\infty \frac{\bar z^{i+1}}{i+1}.
\eeq
By proceeding iteratively we can easily compute the power series in $\bar z$ for any iterated integral to arbitrary power.

In order to obtain compact expressions for our analytic results it is of importance to be able to derive functional relations among our iterated integrals.
One of the big advantages of GPLs is that their functional relations are well studied (see for example~\cite{Duhr:2011zq,Goncharov1998,Panzer:2014caa,Maitre:2005uu,Duhr:2012fh}).
The case of generic iterated integrals is not understood at the same level. 
In ref.~\cite{Remiddi:2017har} it was outlined how relations among iterated integrals involving elliptic functions can be found using IBP identities.
Here, we proceed differently. 

First, note that our final analytic result will be a linear combination of iterated integrals and pre-factors $a_i(\bar z)$,
\beq
\label{eq:coefform}
\sum_i a_i(\bar z) J(\vec \omega_i,\bar z).
\eeq
If there are relations among different iterated integrals appearing in this linear combination then the equation
\beq
\label{eq:relansatz}
\sum_i c_i a_i(\bar z) J(\vec \omega_i,\bar z)=0,\hspace{1cm}, c_i\in \mathbb{Q},
\eeq
can be satisfied for some $c_i\neq 0$ for arbitrary values of $\bar z$. 
The coefficients $a_i(\bar z)$ and corresponding iterated integrals $J(\vec{\omega}_i,\bar z)$ are understood to be identical to those appearing in our final result.
In order to determine the unknown coefficients $c_i$ we expand eq.~\eqref{eq:relansatz} in $\bar z$. 
Every coefficient of every power in $\bar z$ has to vanish separately in order for the equation to be satisfied. 
This allows us to build a system of equations that is large enough to solve for  the unknown coefficients $c_i$.
If we find a certain linear combination of iterated integrals and coefficients that cannot be constrained with this procedure we found a relation of functions.

Let us illustrate the procedure with a trivial example. 
Consider the simple shuffle relation
\beq
c_1 J\left(\frac{1}{1-\bar z},\frac{1}{1+\bar z},\bar z\right)+c_2 J\left(\frac{1}{1+\bar z},\frac{1}{1-\bar z},\bar z\right)+c_3 J\left(\frac{1}{1-\bar z},\bar z\right)J\left(\frac{1}{1+\bar z},\bar z\right)=0,
\eeq
and let us pretend we do not know already know the coefficients $c_i$.
After expanding in $\bar z$ we find
\beq
\frac{7}{60} \left(c_1-c_2\right) \bar{z}^5+\frac{5}{24} \left(c_1+c_2+2 c_3\right) \bar{z}^4+\frac{1}{6} \left(c_1-c_2\right) \bar{z}^3+\frac{1}{2} \left(c_1+c_2+2 c_3\right) \bar{z}^2+\mathcal{O}(\bar z^6)=0
\eeq
We can now create a system of equations by regarding each coefficient in $\bar z$ separately.
\beq
\left(
\begin{array}{cccc}
 \frac{1}{2} & \frac{1}{2} & 1 \\
 \frac{1}{6} & -\frac{1}{6} & 0  \\
 \frac{5}{24} & \frac{5}{24} & \frac{5}{12} \\
 \frac{7}{60} & -\frac{7}{60} & 0 \\
\end{array}
\right).\left(\begin{array}{c} c_1 \\ c_2 \\ c_3 \end{array}\right)=0.
\eeq
Technically, we want to find the kernel of the system of equations. We find that the kernel for our example is spanned by the vector $\{c_1,c_1,-c_1\}^T$. This means we found the shuffle identity
\beq
J\left(\frac{1}{1-\bar{z}},\frac{1}{\bar{z}+1},\bar{z}\right)+J\left(\frac{1}{\bar{z}+1},\frac{1}{1-\bar{z}},\bar{z}\right)-J\left(\frac{1}{1-\bar{z}},\bar z\right) J\left(\frac{1}{\bar{z}+1},\bar{z}\right)=0.
\eeq
Of course this procedure only guarantees that the so-found relations are satisfied up to the order in $\bar z$ at which we truncate our power series. 
However, we may convince ourselves that the relations are correct by computing as many higher order terms as are to our liking.
%
%Applying this procedure to our coefficient functions is still fairly cumbersome since they contain several thousands of iterated integrals. 
%This would imply that we need to derive at least as many terms in the expansion in $\bar z$ to derive the desired relations. 
%The procedure can however be facilitated, by looking at functions that involve only subsets of the letters in our alphabet. 
%For example, we may exclude every function that is purely given in terms of GPLs.
%Furthermore, the procedure is greatly facilitated by the following observation. 
%Relations among iterated integrals often involve only iterated integrals with identical right-most letters in their words. 
%This means we can look at a subset of terms in our coefficient functions that have an identical letter $\omega_1(\bar z)$ but differ in all other letters. 
%This procedure can be extended such that more than one of the right-most letters in a iterated integral are taken to be identical. 
A more involved example of such an identity is given by
\bea
J\left(t_{11},\frac{t_{12}}{1-\bar{z}},\frac{1}{1-\bar{z}}\right)&=& J\left(t_{12},\frac{t_{11}}{1-\bar{z}},\frac{1}{1-\bar{z}}\right)-J\left(t_{21},\frac{t_{12}}{1-\bar{z}},\frac{1}{1-\bar{z}}\right)\nonumber\\
&-&\frac{11}{5} J\left(\frac{t_{21}}{1-\bar{z}},\frac{t_{12}}{1-\bar{z}},\frac{1}{1-\bar{z}}\right)+J\left(t_{22},\frac{t_{11}}{1-\bar{z}},\frac{1}{1-\bar{z}}\right)\nonumber\\
&+&\frac{11}{5} J\left(\frac{t_{22}}{1-\bar{z}},\frac{t_{11}}{1-\bar{z}},\frac{1}{1-\bar{z}}\right)+\frac{1}{5} \left(5 \bar{z}-16\right) t_{11} J\left(\frac{t_{12}}{1-\bar{z}},\frac{1}{1-\bar{z}}\right)\nonumber\\
&-&\frac{1}{5} \left(5 \bar{z}-16\right) t_{12} J\left(\frac{t_{11}}{1-\bar{z}},\frac{1}{1-\bar{z}}\right).
\eea

\subsection{Analytic Solution for Partonic Coefficient Functions}
\label{sec:analyticres}
In the previous sections we described how we derive differential equations for all master integrals required for RRR and RRV partonic coefficient functions.
Furthermore, we outlined how we find a suitable transformation matrix that transforms the differential equations into the form of eq.~\eqref{eq:canform}.
Once, this form is obtained the solution to the differential equations can be conveniently written as in equation eq.~\eqref{eq:diffsol}. 
Iterated integrals as given in eq.~\eqref{eq:iidef} are particularly suited to represent this solution.
Once we calculated all master integrals and computed all boundary conditions we simply insert the master integrals into our IBP reduced matrix elements and obtain the desired result for the partonic coefficient functions.
In this section we describe the structure of our final result for the partonic coefficient functions.

The set of all letters, the so-called alphabet, that appear in the iterated integrals that constitute the Higgs boson cross section at N$^3$LO is given by
\bea
&&\Big\{1,\frac{1}{1-z},\frac{1}{z},\frac{1}{z+1},\frac{1}{\sqrt{z}},\frac{1}{\sqrt{4-z} \sqrt{z}},\frac{\sqrt{z}}{1-z},\frac{1}{\sqrt{z} \sqrt{z+4}},\frac{\sqrt{z}}{\sqrt{z+4}},\frac{1}{\sqrt{4 z+1}},\frac{\sqrt{4 z+1}}{z},\nonumber\\
&&t_{11},t_{12},t_{21},t_{22},\frac{t_{11}}{1-z},\frac{t_{11}}{z},\frac{t_{11}}{z+1},\frac{t_{12}}{1-z},\frac{t_{12}}{z},\frac{t_{12}}{z+1},\frac{t_{21}}{z},\frac{t_{22}}{z}\Big\}.
\eea
Note, that the alphabet required to describe all our master integrals individually contains additional letters that drop out in the final expression. 
%At this point we would like to make a comment on the letters of our alphabet. 
%Singular points of our cross section at $z=1$ or $z=0$ may be associated with physical singularities associated with the production of a Higgs boson or the failure of the Higgs boson mass as a regulator of IR singularities as the partonic center of mass energy becomes infinitely larger than the mass. 
%Other pseudo-thresholds at $m_h^2=-4s$ could be interpreted as thresholds for the creation of particles on a different Riemann sheet. 
%It is much harder to find a physical interpretation for the position of the regular singular points $z=\frac{1}{2} \left(11\pm5 \sqrt{5}\right)$ of the elliptic differential equations.
%We find the presence of these special points as well as the appearance of elliptic functions in the physical cross section intriguing but cannot offer any further interpretation.

The partonic coefficient functions are comprised of iterated integrals with up to five letters. 
Typically we find that there are several thousand different iterated integrals in each partonic coefficient function. 
Applying the procedure outlined in the previous section to find functional identities among these integrals we find that we can express them in terms of only 365 different iterated integrals that cannot be re-written as GPLs in a straight forward fashion. 
Out of those 188 have letters containing elliptic integrals $t_{ij}$. For the remaining ones a representation in terms of GPLs may exist.

Having derived 	moderately compact expressions for our coefficient functions we want to find a method to evaluate them numerically. 
The conceptually simplest way to evaluate the iterated integrals is to perform every integral numerically. 
The fact that all our integrals are real valued and are finite renders this approach straight forward.  
Integrating 5 dimensional integrals numerically is however not particularly fast if a certain level of precision is desired. 
As an alternative, we want to represent the entire partonic coefficient functions in terms of power series expansions.

Let us first investigate for which values of $z$ we can perform a convergent series expansion.
In order to extract this information we regard all singularities and branch points that occur in our alphabet and the algebraic factors of our coefficient functions.
We find that they are located at values of $z$ of
\bea
&&\left\{\frac{1}{2} \left(11+5 \sqrt{5}\right) , 4,1,0 ,\frac{1}{2} \left(11-5 \sqrt{5}\right),-\frac{1}{4},-1,-4\right\}\nonumber\\
&&\sim\left\{11.0902, 4,1,0,-0.0901699,-\frac{1}{4},-1,-4\right\}.
\eea
Here, we included the regular singular points of the differential equations of the elliptic sector, eq.~\eqref{eq:elldiff2}. 
In order to evaluate our functions to high precision within the physical interval, $z\in [0,1]$, we decide to perform a power series expansion around the points $z=1$, $z=\frac{1}{2}$ and $z=0$.
The associated radii of convergence are then  $r_1=1$, $r_{\frac{1}{2}}=\frac{1}{2}$ and  $r_0=\left|\frac{1}{2} \left(11-5 \sqrt{5}\right)\right|$.

To obtain a series expansion around our three different expansion points we perform an expansion of all iterated integrals as outlined in the previous section. 
As the default upper bound for our iterated integrals is the parameter $\bar z$ the expansion around the point $z=1$ can be carried out simply by expanding the iterated integrals at the integrand level and integrating subsequently as demonstrated in eq.~\eqref{eq:serex}. 
In order to obtain an expansion around $z=0$ and $z=\frac{1}{2}$ we first re-express our iterated integrals in terms of iterated integrals with upper integration bound $z$ and $1/2-z$ respectively. 
As outlined in section~\ref{sec:IIs} this procedure requires us to determine certain integration constants which we obtain numerically by matching series expansions around different expansion points. 
To ensure that the numerical error introduced by truncating series expansions is sufficiently small we estimate it as explained in eq.~\eqref{eq:restdef}. 

We expand the coefficient of every iterated integral in the partonic coefficient function separately around each of our three expansion points and combine the result with the expansion for the iterated integrals.
In order to obtain numerical values for the coefficient functions within the unit interval of $z$ we evaluate the expansion around $z=1$ in the interval $z\in[0.75,1]$, the expansion around $z=\frac{1}{2}$ in the interval $z\in \big[\frac{1}{13},0.75\big)$ and the expansion around $z=0$ in the interval $z\in \big[0,\frac{1}{13}\big)$. We truncate the expansion around $z=1$ at $\mathcal{O}((1-z)^{50})$, the expansion around $z=\frac{1}{2}$ at  $\mathcal{O}\left((\frac{1}{2}-z)^{200}\right)$ and the expansion around $z=0$ at  $\mathcal{O}\left(z^{100}\right)$. Using the estimator introduced in eq.~\eqref{eq:restdef} we find that this approximates the coefficient functions at any point in the unit interval to a relative numerical precision of $10^{-10}$ or better. 
This is supported by evaluating the different expansions for several points within the overlaps of their respective domains of convergence and calculating their difference. 
The numerical precision may of course be improved arbitrarily by simply including more terms in the respective series expansions.

\section{Results}
\label{sec:res}
In the previous sections we calculated analytic results for the partonic coefficient functions $\eta_{ij}^{(3)}(z)$. 
Our analytic results agree with the power series around $z=1$ for the same functions obtained in refs.~\cite{Anastasiou:2015ema,Anastasiou:2016cez}. 
The leading behaviour of the coefficient functions as $z\rightarrow 0$ was correctly predicted in ref.~\cite{Hautmann:2002tu}. 
The coefficient function $\eta_{qQ}^{(3)}$ was calculated already in ref.~\cite{Anzai:2015wma} and agrees with our result.
We derived a representation of the coefficient functions in terms of power series expansions that is particularly useful for numerical evaluation. 
In this section we present numerical results for the Higgs boson production cross section through N$^3$LO.

Let us start by regarding the functional dependence of our coefficient functions.
\begin{figure*}[!ht]
\centering
\includegraphics[width=0.9\textwidth]{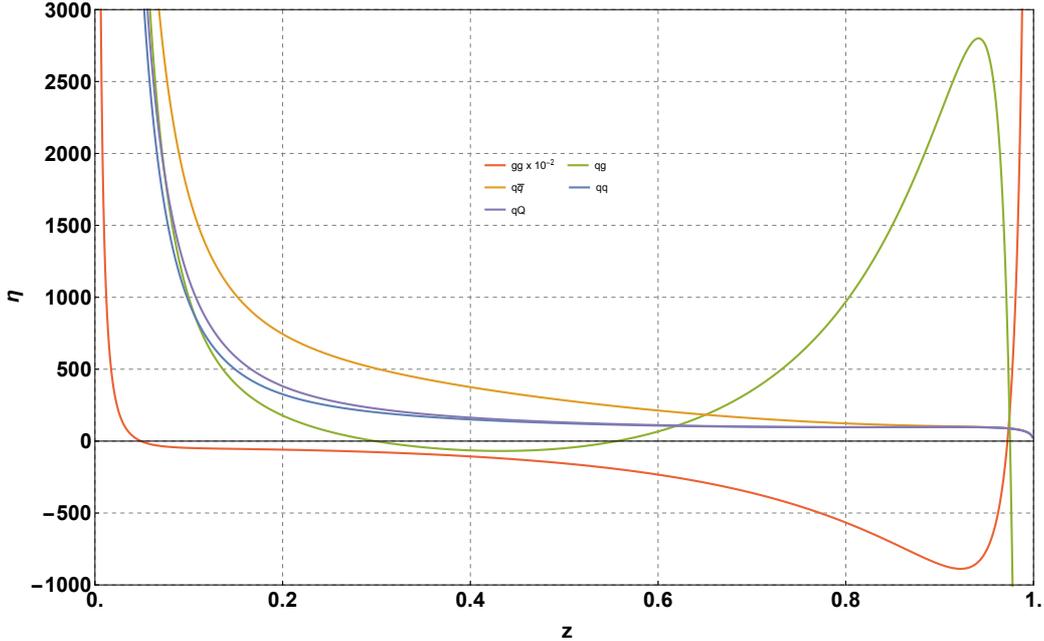}
\caption{\label{fig:coeffunct}
The figure displays the regular N$^3$LO coefficient function for the Higgs boson production cross section for the gg (red), qg (green), q$\bar q$ (orange), qq(blue) and qQ(purple) initial state as a function of the parameter z.
The gg coefficient function was rescaled uniformly by a factor of $10^{-2}$.
}
\end{figure*}
In figure~\ref{fig:coeffunct} we display the shape of the regular coefficient functions for each distinct partonic initial state. 
The quark - gluon and and gluon-gluon initial state coefficient functions behave as $\sim \log^5(1-z)$ as we approach the value $z=1$. 
The coefficient functions with two quarks in the initial state are tending towards zero in this limit. 
The limit $z\rightarrow 0$ is characterised by a power divergence and all coefficient functions behave as $\sim\frac{\log^5(z)}{z}$.

In order to derive physical predictions for hadron collider phenomenology we need to convolute our partonic coefficient functions with parton distribution functions (PDF).
Throughout this article we will use the PDF sets PDF4LHC15~\cite{Rojo:2016ymp}. 
We choose a Higgs boson mass of $125$ GeV and a top quark mass of $m_t(m_t)=162.7$ GeV. We choose a value for the strong coupling constant of $\alpha_S(m_Z=91.1876\text{ GeV})=0.118$.
If not stated otherwise we derive numerical predictions for proton-proton collider with a center of mass energy of $13$ TeV.
We use a private c++ code to perform the numerical convolutions of PDFs and partonic coefficient functions. 

\begin{figure*}[!ht]
\centering
\includegraphics[width=0.92\textwidth]{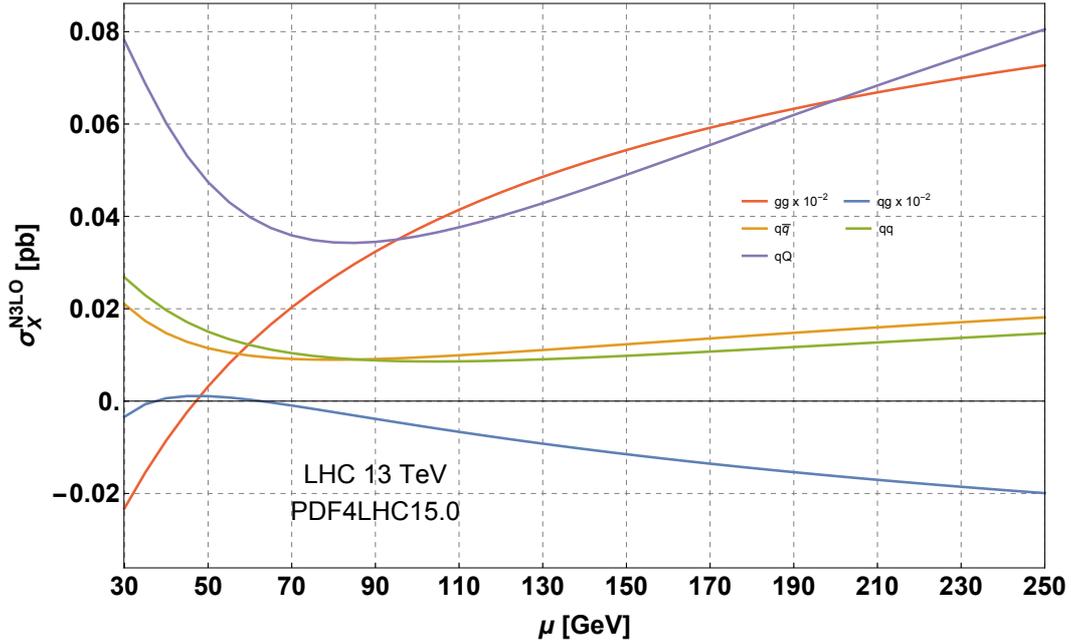}
\caption{\label{fig:variationchan}
The figure displays the contribution of N$^3$LO coefficient function to the Higgs boson production cross section for the gg (red), qg (green), q$\bar q$ (orange), qq(blue) and qQ(purple) initial state as a function of the perturbative scale $\mu$.
The gg and qg coefficient function were rescaled uniformly by a factor of $10^{-2}$.
}
\end{figure*}

In figure~\ref{fig:variationchan} we display the impact of the N$^3$LO corrections on the hadronic cross section for different initial states as a function of the perturbative scale $\mu$. 
The gluon-gluon (red) and quark-gluon (blue) initial state contributions were rescaled by a factor of $10^{-2}$ in order to fit nicely. 
We observe that the numerical impact of these two channels is clearly dominant over all other initial state configurations.  
The nominally smallest corrections for each channel can be found in an interval of $\mu\in [40,90]$ GeV. 

\begin{figure*}[!ht]
\centering
\includegraphics[width=0.92\textwidth]{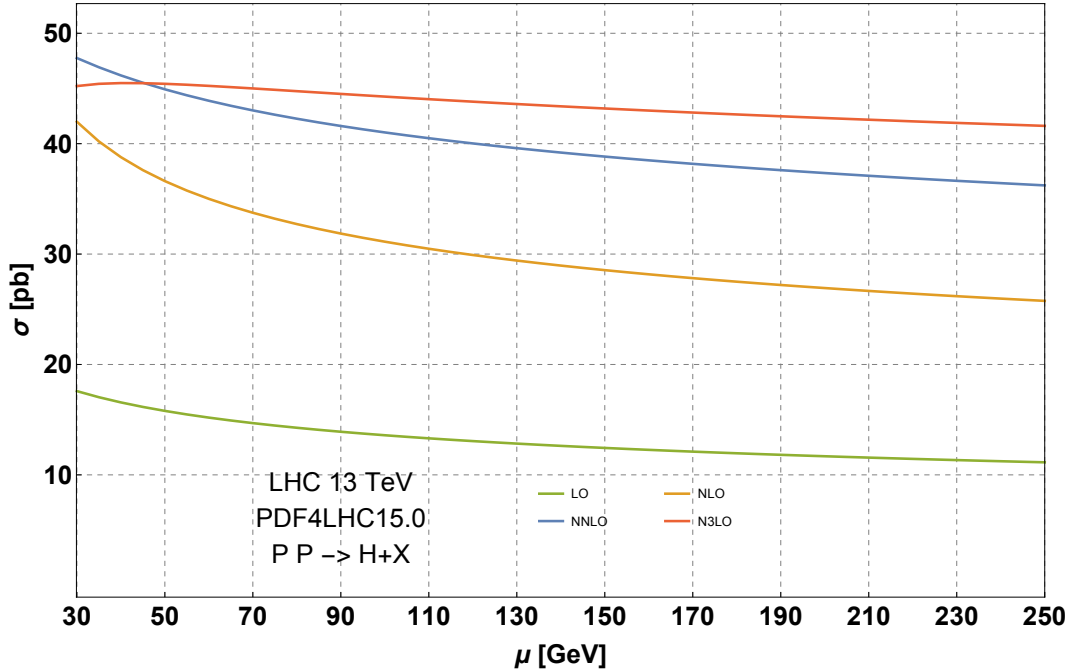}
\caption{\label{fig:variation}
The figure displays the dependence of the Higgs boson production cross section on the perturbative scale $\mu$. 
The green, orange, blue and red lines correspond to a prediction made by truncating the perturbative series at LO, NLO, NNLO and N$^3$LO respectively.
}
\end{figure*}

In figure~\ref{fig:variation} we combine the contribution from all partonic coefficient functions and evaluate their contribution to the hadronic cross section including lower orders in perturbation theory as a function of the  perturbative scale $\mu$.
We show LO, NLO, NNLO and N$^3$LO predictions in green, orange, blue and red respectively. 
We observe that the dependence on the perturbative scale is greatly reduced at N$^3$LO compared to lower orders. 
Furthermore, NNLO and N$^3$LO predictions overlap within the interval of $\mu \in \left[\frac{m_h}{4},m_h\right]$.

To derive a concrete numerical prediction we choose the value of the cross section at $\mu=\frac{m_h}{2}$. 
We vary the perturbative scale in the interval $\left[\frac{m_h}{4},m_h\right]$ in order to estimate the effect of missing higher order corrections at N$^4$LO and beyond.
As can be seen from figure~\ref{fig:variation} this procedure is not conservative enough at leading and next-to-leading order. 
Regarding the progression of the series from NLO onward we observe convergent behaviour. 
The nominal size of the corrections is greatly reduced at each successive order. 
Uncertainty estimates based on scale variation overlap at NNLO and N$^3$LO. 

Our prediction for the Higgs boson production cross section at the LHC based on a computation in perturbative QCD in the large top quark mass limit through N$^3$LO of 
\bea
\label{eq:N3LOPred}
\sigma_{PP\rightarrow H+ X}&=&45.18 \pm{}^{0.31}_{-1.48}\, \text{pb}=45.18 \, \text{pb}  \pm{}^{0.69}_{-3.34} \%.
\eea

\section{Comparison with Results based on a Threshold Expansion}
\label{sec:THcompare}
In ref.~\cite{Anastasiou:2015ema} N$^3$LO corrections to the Higgs boson production cross section were computed using an approximation based on a power series around the point $z=1$ truncated at $\mathcal{O}((1-z)^{30})$.
The expansion around $z=1$ exploits a kinematic enhancement of the gluon luminosity in the collision of protons for lower values of partonic center of mass energy to yield reliable predictions.
The point $z=1$ represents the production threshold for a Higgs boson, i.e. the lowest possible amount of energy required to produce a Higgs boson.
In ref.~\cite{Anastasiou:2016cez}  seven additional terms in the power series were added. 
The quality of a threshold expansion for N$^3$LO corrections was furthermore studied in refs.~\cite{Anastasiou:2014lda,Anastasiou:2014vaa,Herzog:2014wja}.
Having now the complete coefficient functions at our disposal we want to reflect on previous estimates and compare our exact analytical findings to the approximate results.

Using the same set-up as in the previous section to derive numerical predictions we find that the hadronic cross section through N$^3$LO in perturbative QCD in the infinite top quark mass limit based on thirty terms in the threshold limit is given by 
\bea
\sigma_{PP\rightarrow H+ X}^{\text{Threshold-30}}&=&45.07 \pm{}^{0.26}_{-1.43}\, \text{pb}=45.07 \, \text{pb}  \pm{}^{0.58}_{-3.23} \%.
\eea
We observe a difference of $0.11$ pb with respect to our new prediction, eq.~\eqref{eq:N3LOPred}. 
The scale variation interval in eq.~\eqref{eq:N3LOPred} is slightly larger. 
In ref.~\cite{Anastasiou:2016cez} it was estimated the effect of missing higher order terms in the threshold expansion are less than $0.18$ pb. 
We now see that this estimate was sufficiently conservative.

\begin{figure*}[!ht]
\centering
\begin{subfigure}[b]{0.47\textwidth}
\includegraphics[width=0.95\textwidth]{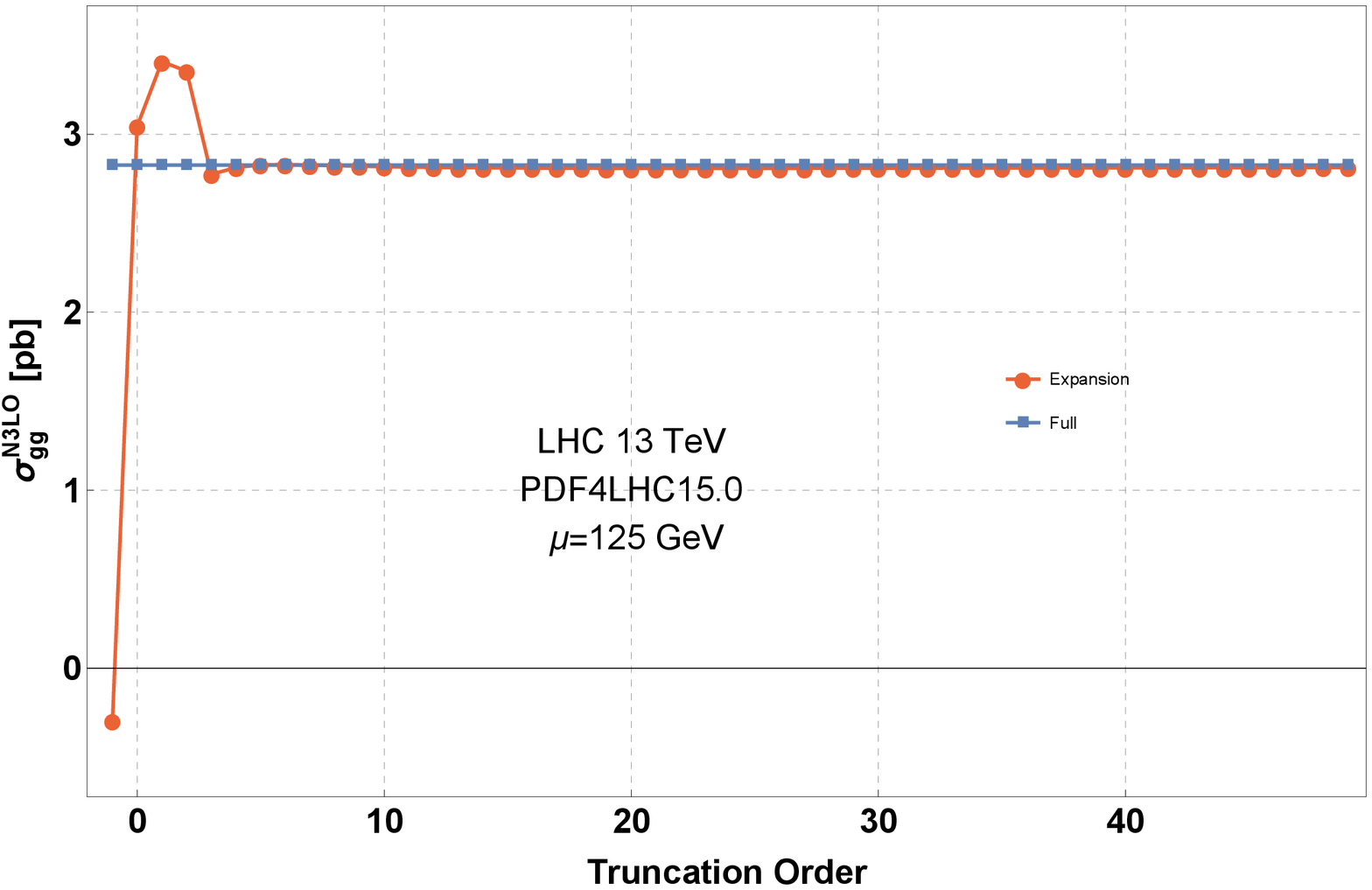}
\caption{ 
}
\end{subfigure}
\begin{subfigure}[b]{0.47\textwidth}
\includegraphics[width=0.95\textwidth]{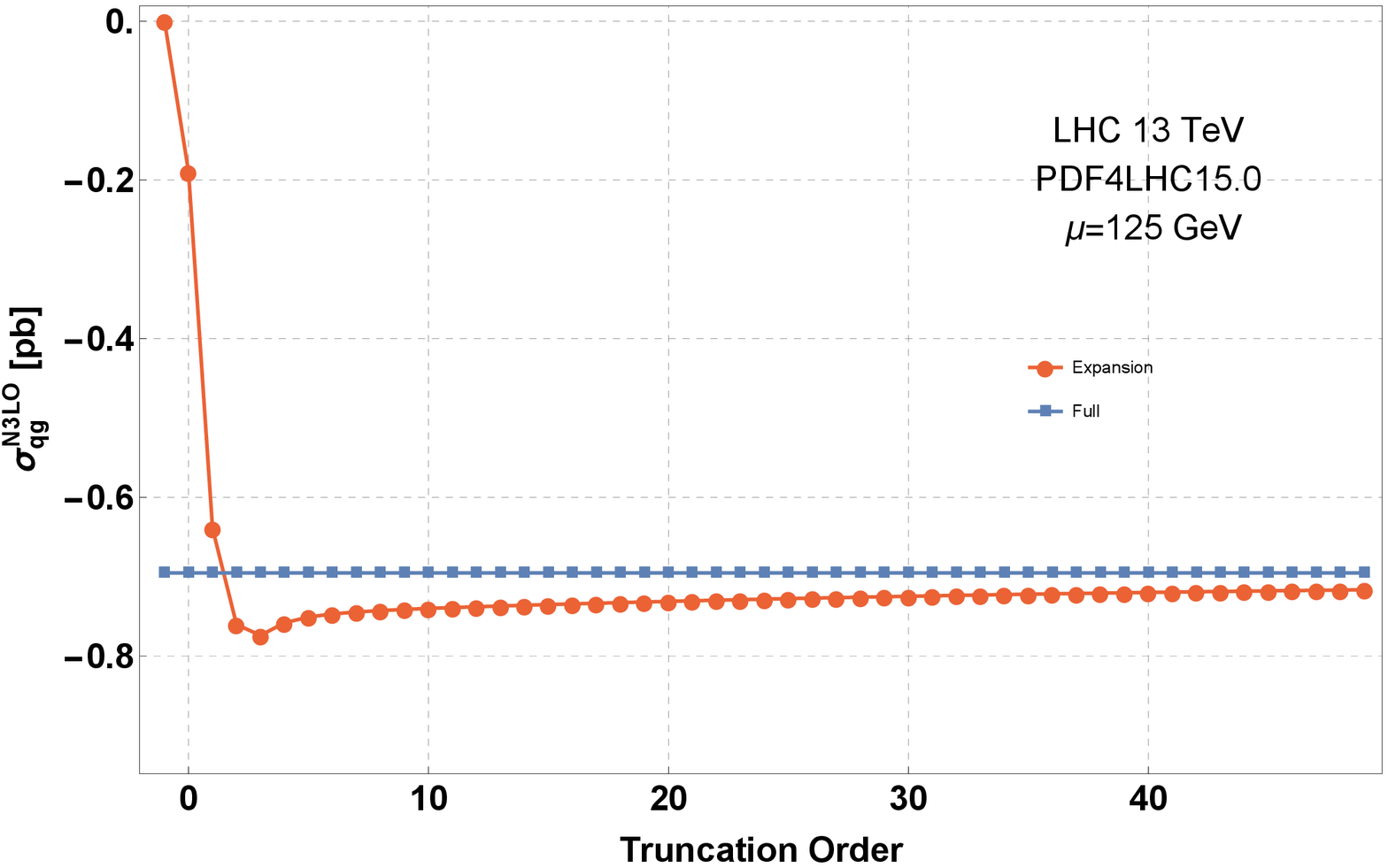}
\caption{ 
}
\end{subfigure}
\begin{subfigure}[b]{0.457\textwidth}
\includegraphics[width=0.95\textwidth]{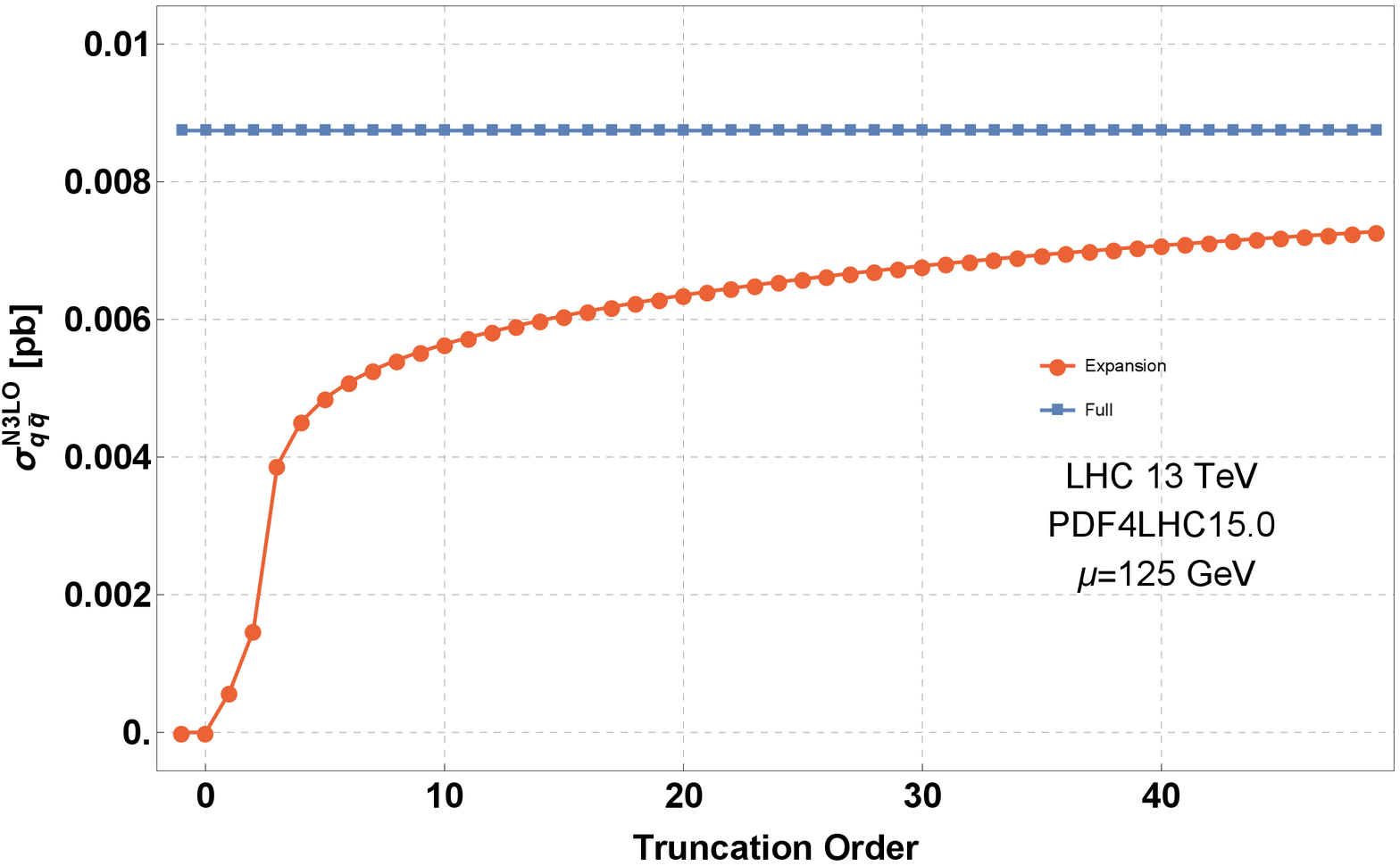}
\caption{ 
}
\end{subfigure}
\begin{subfigure}[b]{0.47\textwidth}
\includegraphics[width=0.95\textwidth]{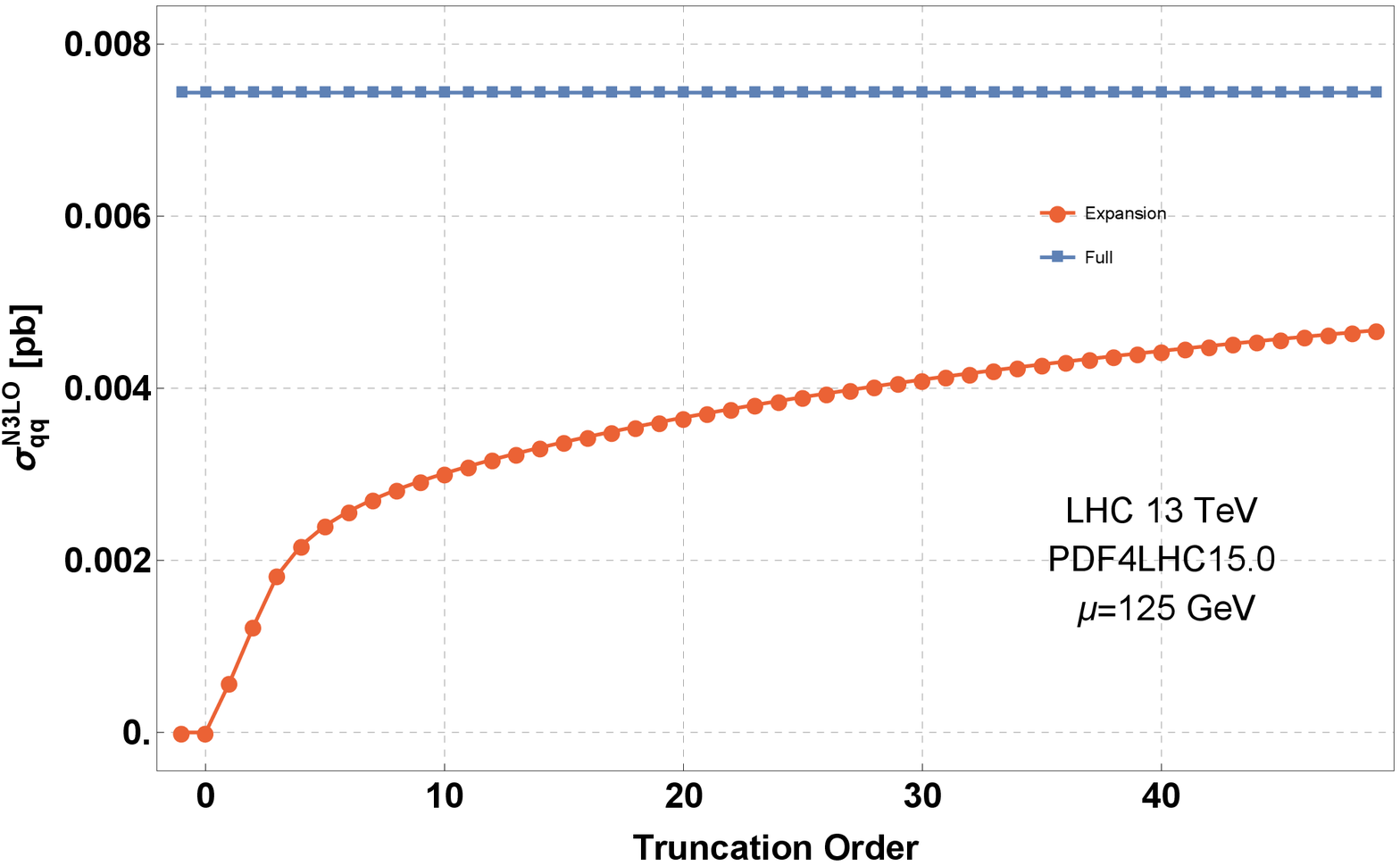}
\caption{ 
}
\end{subfigure}
\begin{subfigure}[b]{0.47\textwidth}
\includegraphics[width=0.95\textwidth]{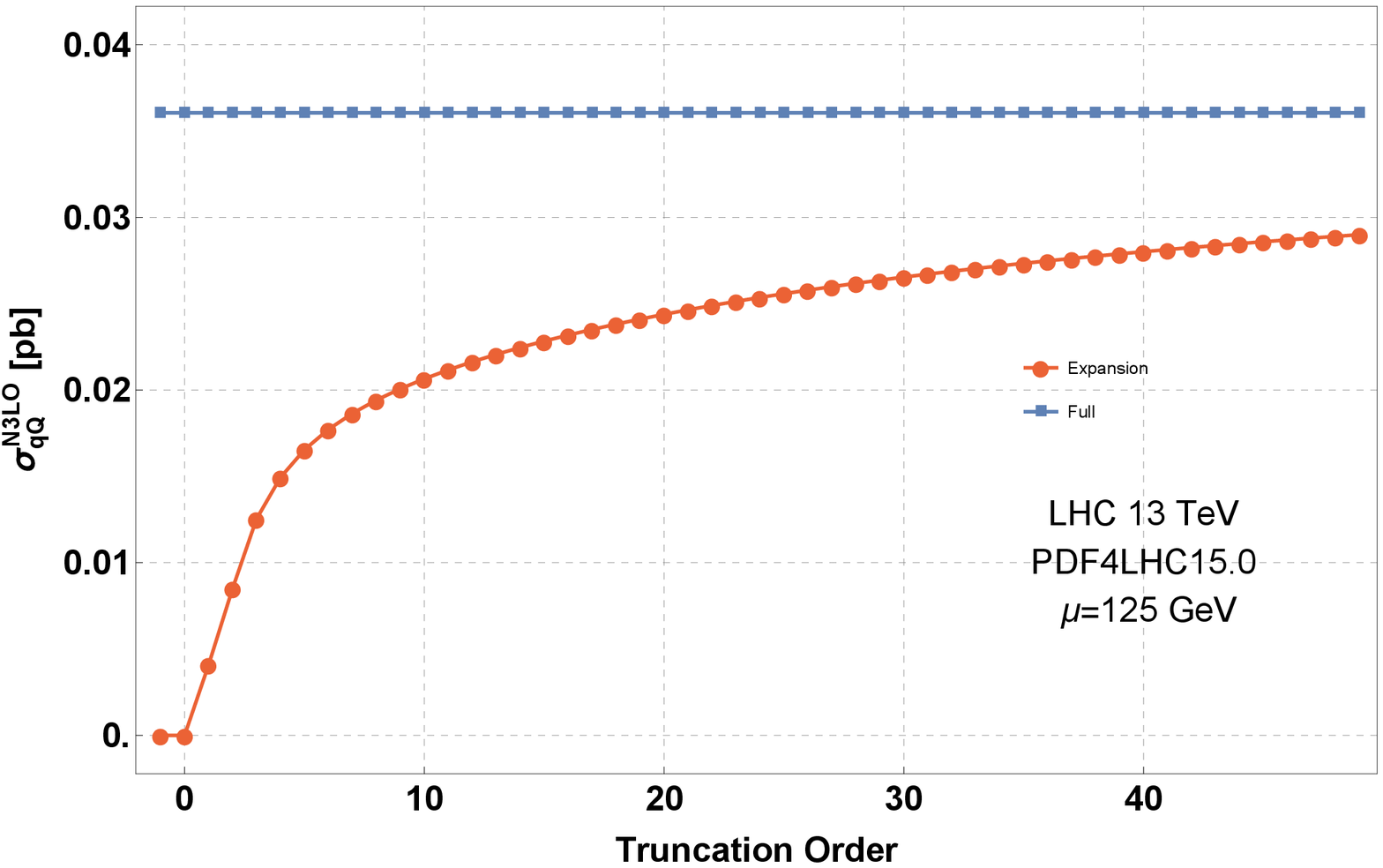}
\caption{ 
}
\end{subfigure}
\begin{subfigure}[b]{0.47\textwidth}
\includegraphics[width=0.95\textwidth]{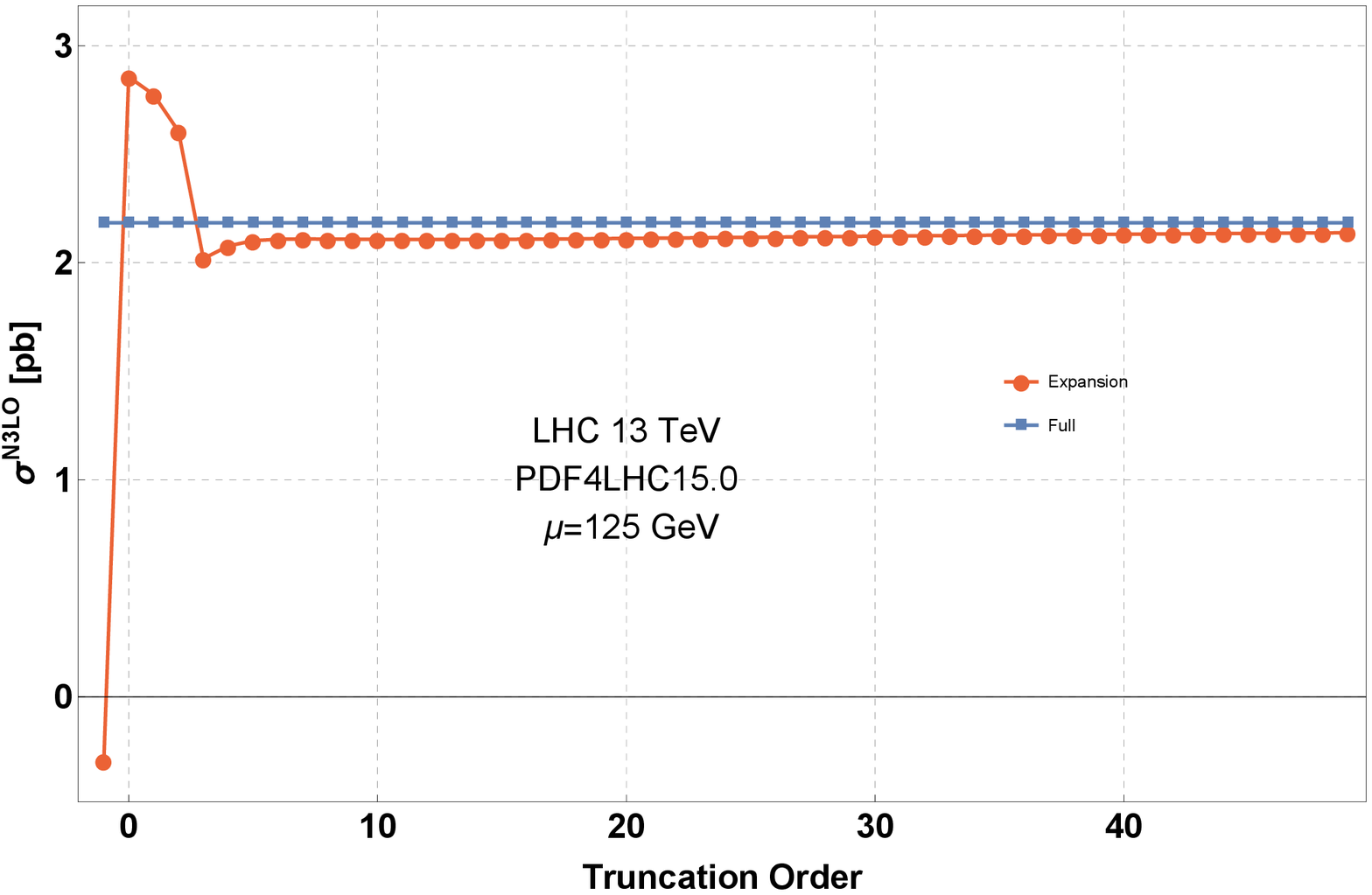}
\caption{ \label{fig:truncall}
}
\end{subfigure}
\caption{\label{fig:trunc}
The figure shows in red the contribution of the partonic coefficient function to the N$^3$LO correction of the Higgs boson cross section approximated by a threshold expansion. 
The x-axis labels the order at which the expansion is truncated. The line in blue represents the contribution to all orders in the threshold expansion and is displayed as a reference.
Figures (a), (b), (c), (d), (e) and (f) show the contribution due to the gg, qg, q$\bar q$, qq, qQ initial state and the sum of all channels respectively.
}
\end{figure*}

In the remainder of this section we want to study the behaviour of N$^3$LO corrections as a function of the order where the threshold expansion is truncated. 
In particular we want to investigate its performance for contributions arising from different partonic initial states. 
In figure~\ref{fig:trunc} we show the N$^3$LO correction due to different initial sate partons based on a threshold expansion (red) as a function of the order at which the expansion is truncated.
In blue we also display our new result to all orders in the threshold expansion as a reference.
We observe that the first four terms show particularly large changes in the derived prediction. 
Starting from the fifth term we observe slow asymptotic improvement towards the full result. 
The nominally largest gluon-gluon and quark-gluon channels are approximated better than their purely quark initiated counter parts. 
The sum of all channels can be seen in figure~\ref{fig:truncall}.

\begin{figure*}[!ht]
\centering
\begin{subfigure}[b]{0.47\textwidth}
\includegraphics[width=0.95\textwidth]{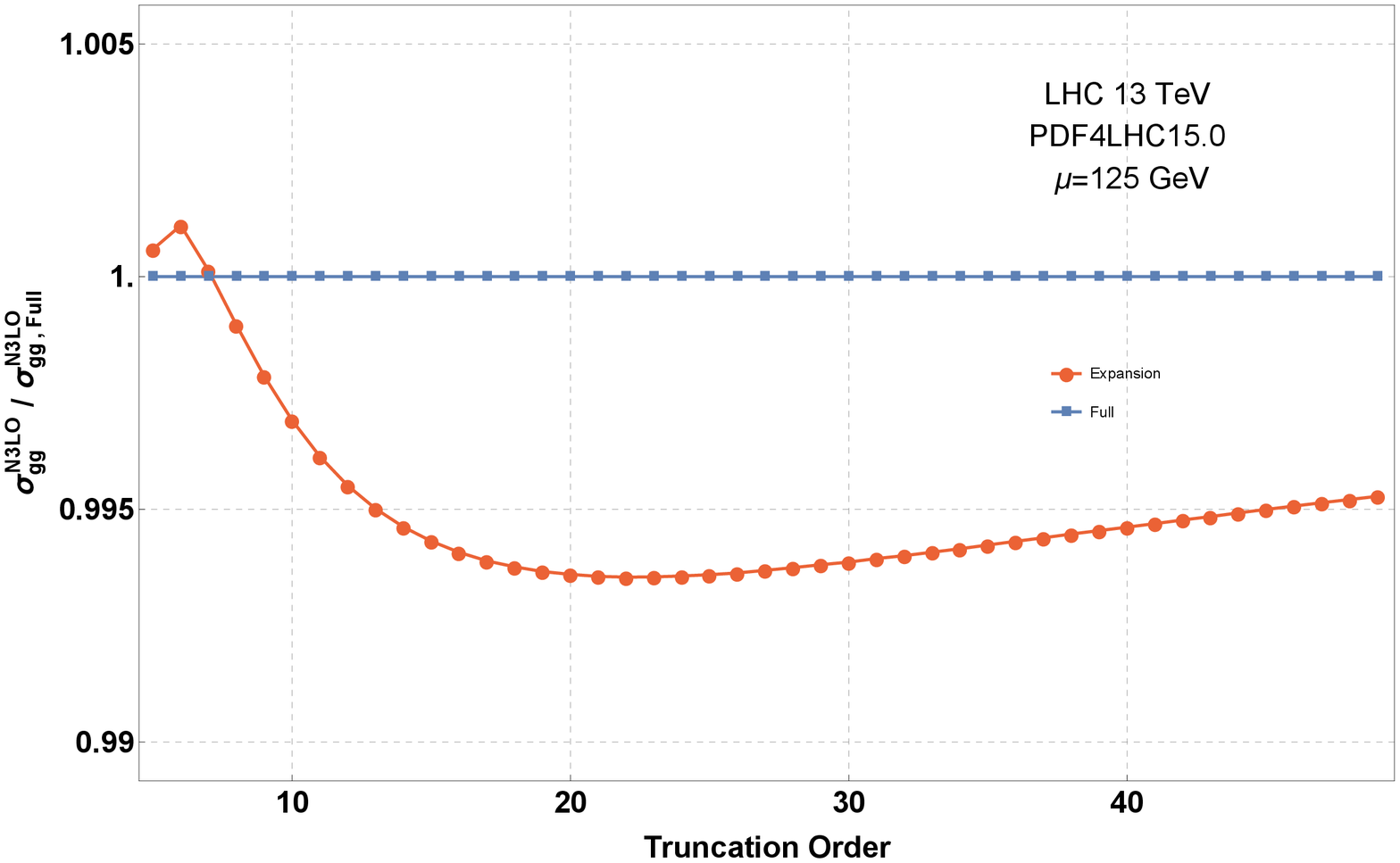}
\caption{ 
}
\end{subfigure}
\begin{subfigure}[b]{0.47\textwidth}
\includegraphics[width=0.95\textwidth]{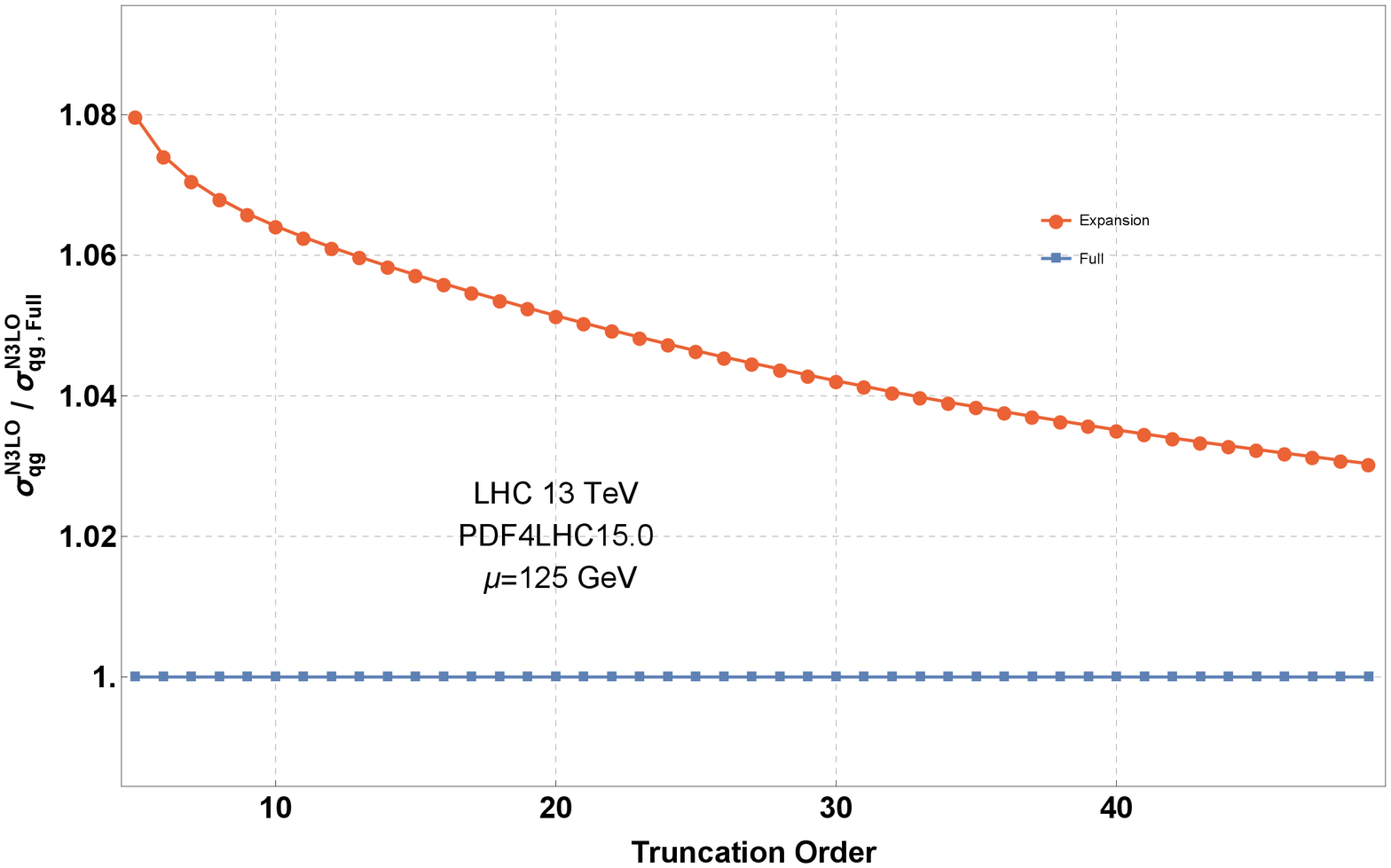}
\caption{ 
}
\end{subfigure}
\begin{subfigure}[b]{0.47\textwidth}
\includegraphics[width=0.95\textwidth]{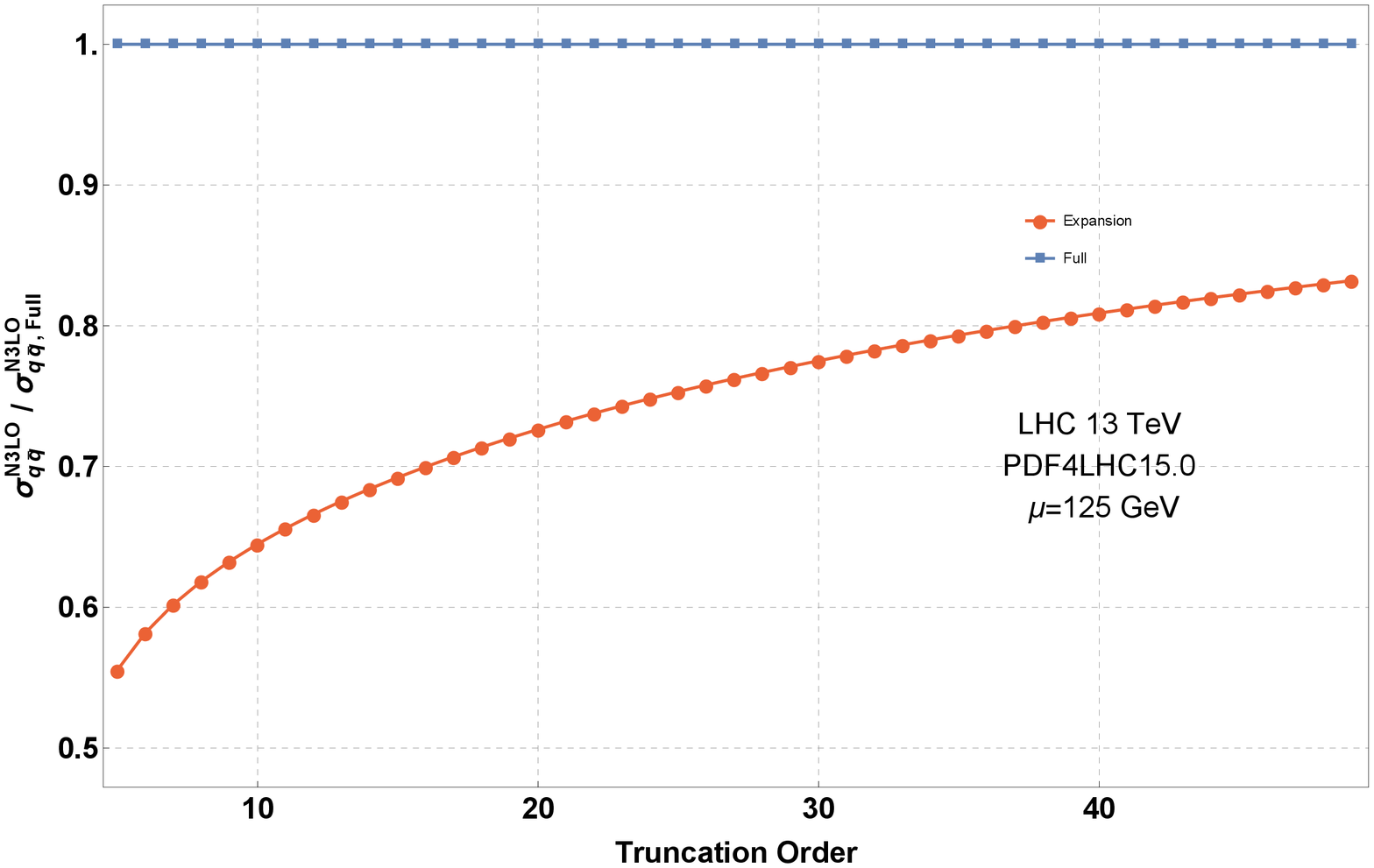}
\caption{ 
}
\end{subfigure}
\begin{subfigure}[b]{0.47\textwidth}
\includegraphics[width=0.95\textwidth]{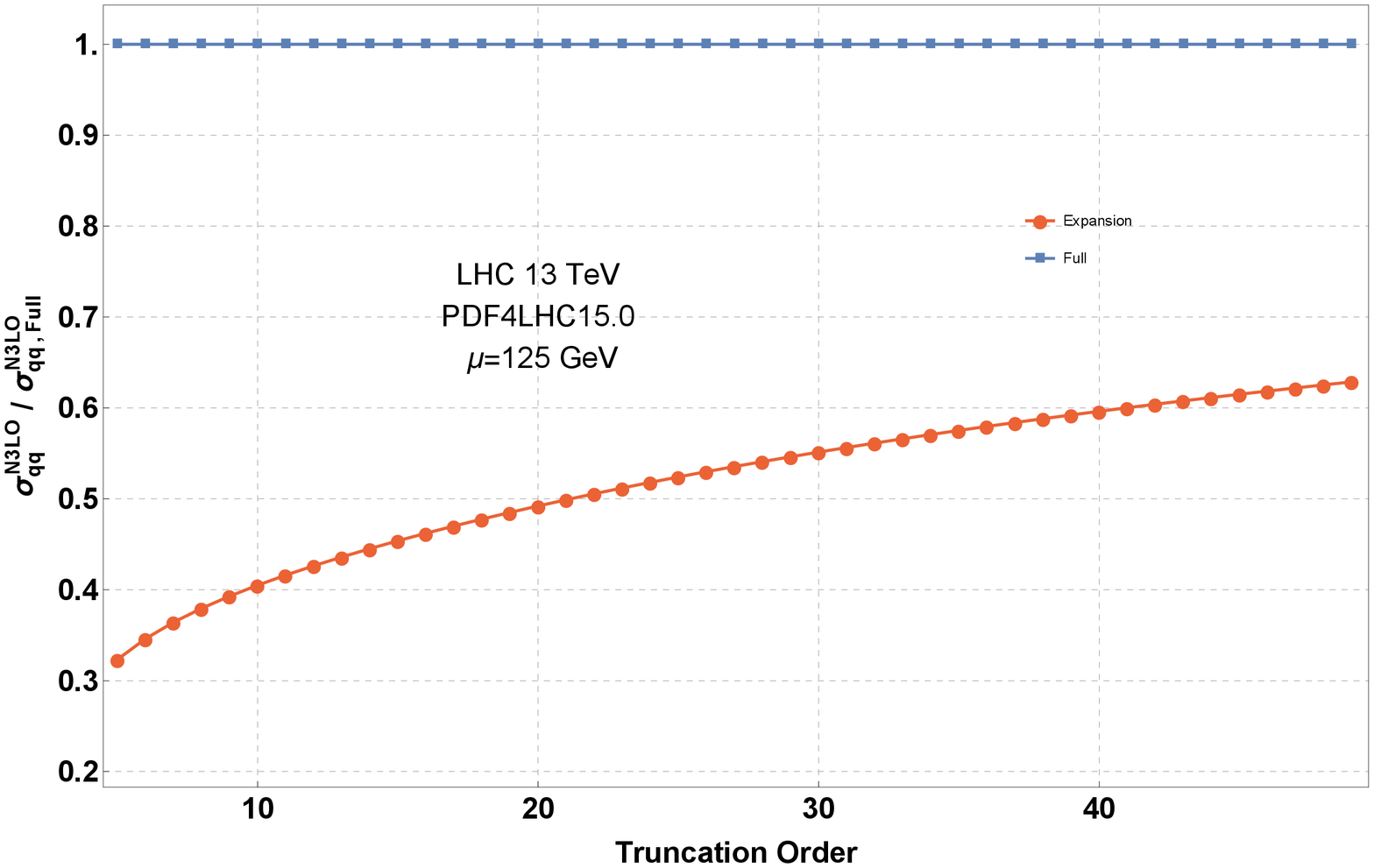}
\caption{ 
}
\end{subfigure}
\begin{subfigure}[b]{0.47\textwidth}
\includegraphics[width=0.95\textwidth]{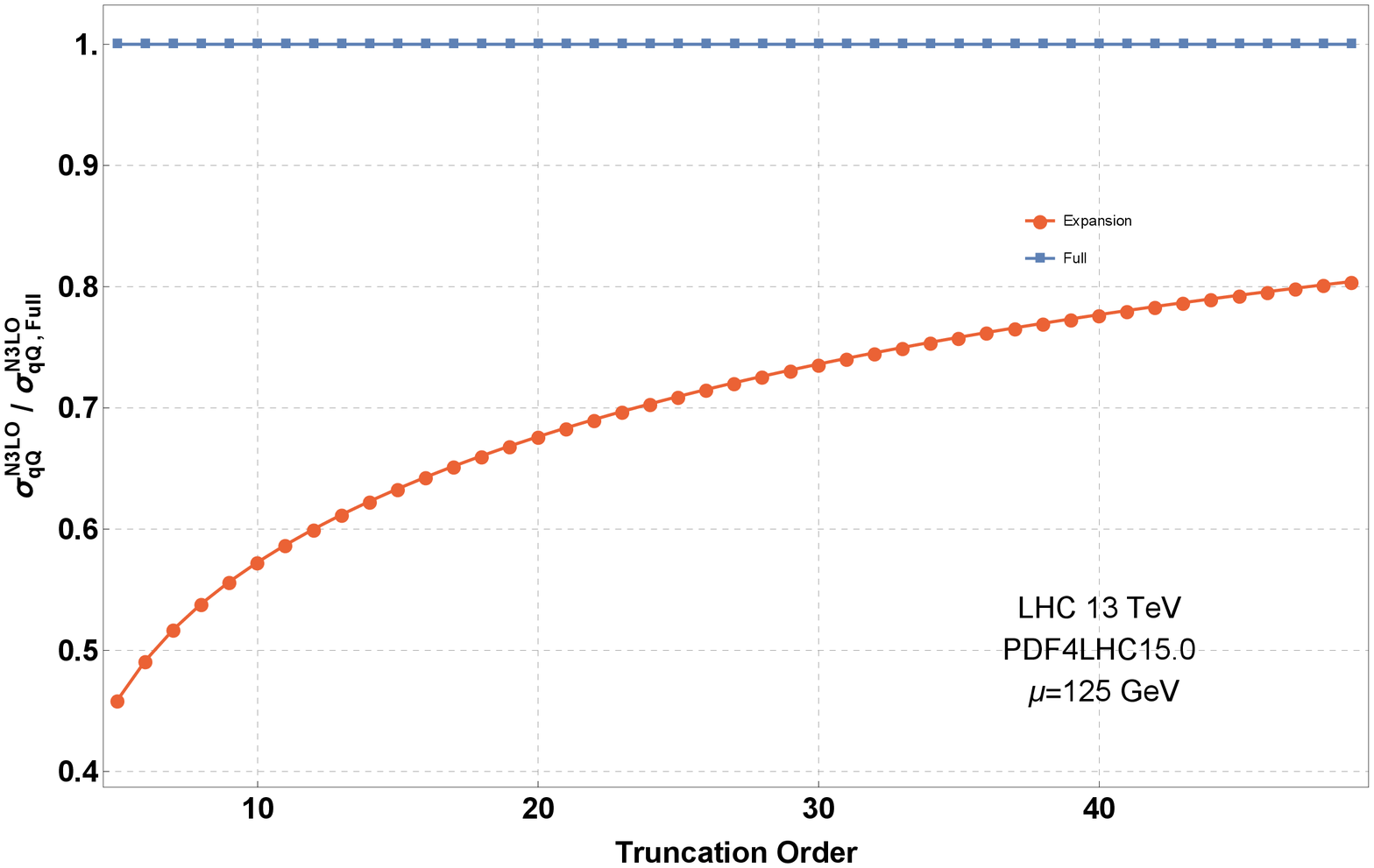}
\caption{ 
}
\end{subfigure}
\begin{subfigure}[b]{0.47\textwidth}
\includegraphics[width=0.95\textwidth]{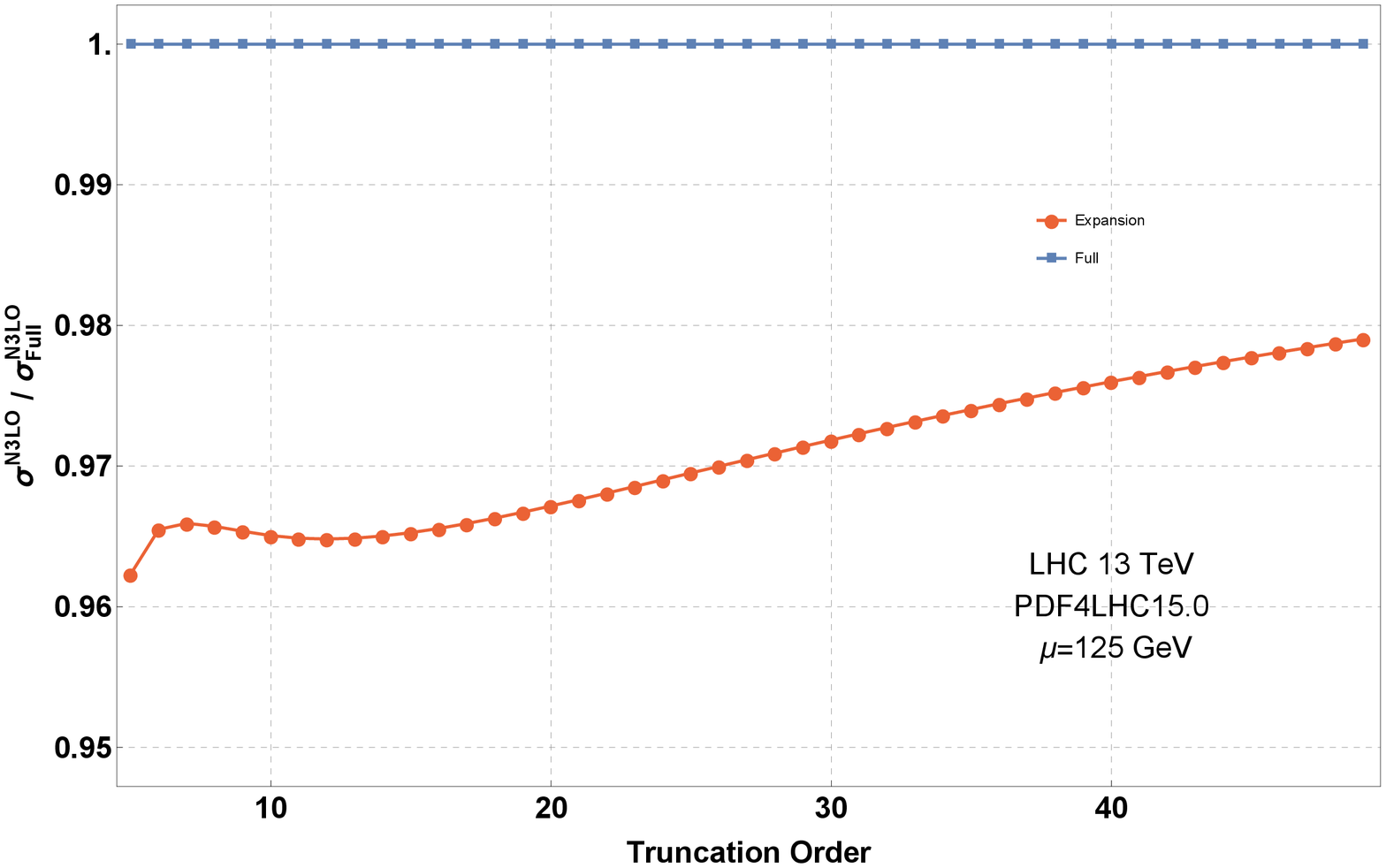}
\caption{ 
}
\end{subfigure}
\caption{\label{fig:truncrel}
The figure shows in red the contribution of the partonic coefficient function to the N$^3$LO correction of the Higgs boson cross section approximated by a threshold expansion normalised to the all order result. 
The x-axis labels the order at which the expansion is truncated. The line in blue represents the contribution to all orders in the threshold expansion and is displayed as a reference.
Figures (a), (b), (c), (d), (e) and (f) show the contribution due to the gg, qg, q$\bar q$, qq, qQ initial state and the sum of all channels respectively.
}
\end{figure*}
In order to see more clearly the quality of the threshold expansion for each channel we show in figure~\ref{fig:truncrel} the impact of N$^3$LO corrections on the hadronic cross section due to different partonic initial states.
The predictions in red are now based on a threshold expansion normalised to the respective all order result. 
The x-axis shows the order at which the threshold expansion is truncated.
The line in blue at one serves as a reference. 
We observe that contributions originating from the gluon-gluon channel are approximated within several per-mille including only a few terms in the expansion. 
Similarly the quark-gluon initiated contributions are approximated reasonably well below a level of ten percent. 
All other contributions are considerably different from the exact result and receive corrections of the order of 100 \% even with thirty terms in the expansion.
Their nominal effect on the inclusive cross section is however negligible.
The fact that the threshold expansion works best for gluonic initial states can be explained by the fact that the probability to extract gluons from a proton is peaked towards lower momentum fractions, i.e. closer to the production threshold.
For quarks this enhancement is not as large. 
The relatively slow improvement towards the exact result of the predictions as more and more terms in the threshold expansion are included can be understood from the high energy behaviour of the partonic coefficient functions.
As we displayed in fig.~\ref{fig:coeffunct} the coefficient functions have a power like divergence $\sim \frac{\log^5(z)}{z}$ as $z\rightarrow 0$. 
While the threshold expansion is formerly convergent within the entire physical interval a relatively slow convergence to capture the high energy behaviour can be expected.

\section{Conclusions}
\label{sec:conclusions}
In this article we present an exact computation of the Higgs boson production cross section at hadron colliders through N$^3$LO in perturbative QCD in the infinite top quark mass limit.
The main result of this article are analytic formulae for N$^3$LO corrections to the regular partonic coefficient functions. 
We provide these functions in an ancillary file together with the arXiv submission of this article.

To obtain our result we compute matrix elements for the production of a Higgs boson in association with three partons at tree level and with two partons at the one-loop level.
In order to perform required phase space integrals we employ the framework of reverse unitarity and make use of loop integration techniques such as IBP identities and master integrals.
We compute all required master integrals using the framework of differential equations. 
When solving differential equations we encounter elliptic integrals in the solution for triple real radiation master integrals. 
We find that an analytic solution for our master integrals can be easily found by embedding the solution of our differential equations in a fairly general class of iterated integrals.
We discuss in detail how we find relations among iterated integrals involving elliptic functions and how we evaluate them efficiently numerically.

Having obtained analytic expressions for all required partonic cross sections we embed them in a numerical code and derive predictions for hadron collider cross sections.
We find that N$^3$LO corrections are small compared to corrections at previous orders and that the dependence on the perturbative scale is greatly reduced.
We perform a detailed comparison with a previous approximation of N$^3$LO corrections based on an expansion around the production threshold of the Higgs boson including 37 terms~\cite{Anastasiou:2015ema,Anastasiou:2016cez}.
We observe that our new results are in excellent agreement with this approximation. 
Dominant contributions due to gluon initiated partonic cross sections are approximated rather well by the threshold expansion. 
Quark initiated contributions on the other hand are approximated rather poorly.
The estimate of missing higher orders in the threshold expansion in refs.~\cite{Anastasiou:2015ema,Anastasiou:2016cez} was sufficiently conservative to cover the difference to the exact result.

\begin{table}[!h]
\normalsize\setlength{\tabcolsep}{2pt}
\begin{center}
    \begin{tabular}{rrrrr}
        \toprule
        \multicolumn{1}{c}{$\textrm{E}_{\textrm{CM}}$}&
        \multicolumn{1}{c}{$\sigma$}&
        \multicolumn{1}{c}{$\delta(\textrm{theory})$}&
        \multicolumn{1}{c}{$\delta(\textrm{PDF})$}&
        \multicolumn{1}{c}{$\delta(\alpha_s)$}\\\midrule
        2 TeV & 1.10 pb & ${}^{+0.05\textrm{pb}}_{-0.09\textrm{pb}}\,\left({}^{+4.17\%}_{-8.02\%}\right) $ & $\pm\,0.03\,\textrm{pb}\,(\pm\,3.17\%)$ & ${}^{+0.04\textrm{pb}}_{-0.04\textrm{pb}}\,\left({}^{+3.69\%}_{-3.36\%}\right) $ \\\midrule
        7 TeV & 16.87 pb & ${}^{+0.70\textrm{pb}}_{-1.14\textrm{pb}}\,\left({}^{+4.17\%}_{-6.76\%}\right) $ & $\pm\,0.31\,\textrm{pb}\,(\pm\,1.89\%)$ & ${}^{+0.44\textrm{pb}}_{-0.45\textrm{pb}}\,\left({}^{+2.66\%}_{-2.68\%}\right) $ \\\midrule
8 TeV & 21.45 pb & ${}^{+0.90\textrm{pb}}_{-1.43\textrm{pb}}\,\left({}^{+4.18\%}_{-6.69\%}\right) $ & $\pm\,0.40\,\textrm{pb}\,(\pm\,1.87\%)$ & ${}^{+0.56\textrm{pb}}_{-0.56\textrm{pb}}\,\left({}^{+2.63\%}_{-2.66\%}\right) $ \\\midrule
13 TeV & 48.68 pb & ${}^{+2.07\textrm{pb}}_{-3.16\textrm{pb}}\,\left({}^{+4.26\%}_{-6.48\%}\right) $ & $\pm\,0.89\,\textrm{pb}\,(\pm\,1.85\%)$ & ${}^{+1.25\textrm{pb}}_{-1.26\textrm{pb}}\,\left({}^{+2.59\%}_{-2.62\%}\right) $ \\\midrule
14 TeV & 54.80 pb & ${}^{+2.34\textrm{pb}}_{-3.54\textrm{pb}}\,\left({}^{+4.28\%}_{-6.46\%}\right) $ & $\pm\,1.00\,\textrm{pb}\,(\pm\,1.86\%)$ & ${}^{+1.40\textrm{pb}}_{-1.42\textrm{pb}}\,\left({}^{+2.60\%}_{-2.62\%}\right) $ \\\midrule
28 TeV & 154.63 pb & ${}^{+7.02\textrm{pb}}_{-9.93\textrm{pb}}\,\left({}^{+4.54\%}_{-6.42\%}\right) $ & $\pm\,2.98\,\textrm{pb}\,(\pm\,1.96\%)$ & ${}^{+4.10\textrm{pb}}_{-4.03\textrm{pb}}\,\left({}^{+2.70\%}_{-2.65\%}\right) $ \\\midrule
100 TeV & 808.23 pb & ${}^{+44.53\textrm{pb}}_{-56.95\textrm{pb}}\,\left({}^{+5.51\%}_{-7.05\%}\right) $ & $\pm\,19.98\,\textrm{pb}\,(\pm\,2.51\%)$ & ${}^{+24.89\textrm{pb}}_{-21.71\textrm{pb}}\,\left({}^{+3.12\%}_{-2.72\%}\right) $ \\\bottomrule
    \end{tabular}
    \caption{Cross sections and uncertainties as function of the collider center of mass energy.\label{tab:xs_relerr}}
\end{center}
\end{table}
To derive precise predictions for hadron collider phenomenology many effects beyond the effective theory cross section considered in this article have to be take into account.
The finiteness of quark masses and neglected electro-weak effects play an important role. 
It is particularly important to critically asses all non-negligible sources of uncertainty. 
A detailed study of the inclusive production cross section for the Higgs boson considering all such effects was conducted in ref.~\cite{Anastasiou:2016cez}. 
Repeating this analysis is beyond the scope of this article. 
However, it easily possible to modify the final predictions for hadron collider cross sections of ref.~\cite{Anastasiou:2016cez} such that the results of this article are taken into account.
Specifically, we include the exact contributions to the cross section at N$^3$LO in the EFT and remove uncertainties due to the truncation of the threshold expansion. 
Otherwise, we can simply use the results of ref.~\cite{Anastasiou:2016cez} that are neatly combined in a new numerical code \texttt{iHixs\,2}~\cite{Dulat2018}. 
In table~\ref{tab:xs_relerr} we show updated predictions for the gluon fusion Higgs boson production cross section at the LHC as in ref.~\cite{Dulat2018}.
%A dedicated numerical code that combines the results of ref.~\cite{Anastasiou:2016cez} with the exact coefficient functions obtained here in order to provide phenomenological predictions for the LHC is part of future work

\section{Acknowledgements}
I would like to thank Babis Anastasiou, Claude Duhr, Falko Dulat and Franz Herzog for many invaluable discussions.
I would like to thank Claude Duhr and Babis Anastasiou for useful comments on the manuscript.
I am grateful to Lorenzo Tancredi and Stefan Weinzierl for useful discussions about elliptic integrals appearing in this computation.
Furthermore, I would like to thank Achilleas Lazopoulos and Falko Dulat for comparisons of numerical results using the latest version of $\texttt{iHixs\,2}$~\cite{Dulat2018}.
My research was supported by the European Comission through the ERC Consolidator Grant HICCUP (No. 614577).

\appendix
\section{The Elliptic Integral}
\label{sec:APPEll}

In section~\ref{sec:elliptic} we discuss a coupled system of two differential equations that describes the homogeneous solution to master integrals appearing in triple real radiation matrix elements when integrated over phase space.
The particular system is given by
\beq
\frac{\partial}{\partial z} \left(\begin{array}{c} E_4^0 \\ E_1^0 \end{array}\right)=
\left(
\begin{array}{cc}
 0 &\frac{1}{z}   \\
\frac{3-z}{z^2-11 z-1}  &  \frac{11-2 z}{z^2-11 z-1}\\
\end{array}
\right).
 \left(\begin{array}{c} E_4^0 \\ E_1^0 \end{array}\right).
\eeq
Equivalently, we can say that $E_4^0$ satisfies a second order differential equation.
\bea
\label{eq:secondell}
&&\frac{\partial^2}{\partial z^2} E_4^0+\frac{\left(3 z^2-22 z-1\right) }{z \left(z^2-11 z-1\right)} \frac{\partial}{\partial z}E_4^0+\frac{(z-3)}{z \left(z^2-11 z-1\right)}E_4^0=0.\nonumber\\
&&E_1^0=z\frac{\partial}{\partial z}E_4^0.
\eea
First, a solution to this differential equation was found by Stefan Weinzierl in terms of an elliptic integral.

%Elliptic integrals have made their appearance in several cases in particle physics (see for example refs.~\cite{Bonciani:2016qxi,Laporta:2004rb,Remiddi:2016gno}) and represent a significant challenge to obtain analytic results for multi-loop integrals. 
The homogeneous part of a differential equation for a Feynman integral has to be satisfied by the maximum cut of the corresponding Feynman integral. 
In ref.~\cite{Henn2013} 
it was proposed that it is sufficient to normalise the leading singularities of Feynman integrals to constants in order to decouple their differential equations order by order in the dimensional regulator.
For this to hold true the physical linear combinations of leading singularities themselves must satisfy the homogeneous differential equation for $\epsilon=0$.
Computing the leading singularity of $E_4$ we find
\beq
\text {Leading Singularity }( E_4)\sim\int dx \frac{\theta\left((x-z) \left(x^3-x^2 z+2 x^2+2 x z+x-z\right)\right)}{\sqrt{(x-z) \left(x^3-x^2 z+2 x^2+2 x z+x-z\right)}}.
\eeq
We can rewrite the quartic polynomial under the square root as 
\beq
(x-z) \left(x^3-x^2 z+2 x^2+2 x z+x-z\right)=(x-r_1)(x-r_2)(x-r_3)(x-r_4).
\eeq
Following the prescription of ref.~\cite{Laporta:2004rb} we define two integrals 
\bea
I_1&=&\int_{r_2}^{r_3} dx \frac{1}{\sqrt{(x-r_1)(x-r_2)(x-r_3)(x-r_4)}}\nonumber\\
&=&\frac{2}{\sqrt{(r_4-r_2)(r_3-r_1)}} K(1-m).\nonumber\\
I_2&=&\int_{r_3}^{r_4} dx \frac{1}{\sqrt{(x-r_1)(x-r_2)(x-r_3)(x-r_4)}}\nonumber\\
&=&\frac{2}{\sqrt{(r_4-r_2)(r_3-r_1)}} K(m).\nonumber
\eea
Here, $K(m)$ is the complete elliptic integral of the first kind. 
We find that both integrals $I_1$ and $I_2$ are solutions to our second order differential equation eq.~\ref{eq:secondell}.
In principle we could now follow a procedure outlined in ref.~\cite{Laporta:2004rb} to construct a transformation matrix $T_E$ that allows us to decouple the system of differential equations order by order in $\epsilon$.
Specifically, we find that the functions $t_{ij}(z)$ defined in section~\ref{sec:elliptic} are given by linear combinations
\beq
t_{ij}(z)=c_1 I_1+c_2 I_2 +c_3 z\frac{\partial}{\partial z}I_1+c_4 z\frac{\partial}{\partial z}I_2,\hspace{1cm} c_i\in \mathbb{C}.
\eeq
The derivatives of the functions $I_1$ and $I_2$ with respect to z yield a sum of elliptic integrals of first and second kind with algebraic pre-factors. 
We can determine the coefficients $c_i$ analytically by equating the power series expansions of the above equation with the results obtained in section~\ref{sec:elliptic}. 
However, any of these analytic expressions is quite unwieldy.

\section{Various Ingredients for Higgs Boson Production}
\label{sec:consts}
In this appendix we summarise various standard ingredients for the perturbative calculation of the inclusive Higgs boson production cross section.

In order to perform renormalisation in the $\overline{\text{MS}}$ scheme we substitute the bare coupling and Wilson coefficient as 
\bea
\alpha_S^0&=&\alpha_S(\mu^2) \left(\frac{\mu^2}{4\pi}\right)^{\epsilon} e^{\epsilon \gamma_E} Z_\alpha.\nonumber\\
C^0&=&C Z_C.
\eea
The renormalisation factors for the strong coupling constant and Wilson coefficient required for a computation through N$^3$LO~\cite{Gehrmann2010} are given by
\bea
Z_\alpha&=&1+\frac{\alpha_S}{\pi}\left(-\frac{\beta_0}{\epsilon }\right)+\left(\frac{\alpha_S}{\pi}\right)^2\left(\frac{\beta_0^2}{\epsilon ^2}-\frac{\beta_1}{2 \epsilon }\right)+\left(\frac{\alpha_S}{\pi}\right)^3\left(-\frac{\beta_0^3}{\epsilon ^3}+\frac{7 \beta_1 \beta_0}{6 \epsilon ^2}-\frac{\beta_2}{3 \epsilon }\right)+\mathcal{O}(\alpha_S^4).\nonumber\\
Z_C&=&1-\frac{\alpha_S}{\pi}\left(\frac{\beta_0}{\epsilon}\right) +\left(\frac{\alpha_S}{\pi}\right)^2\left(\frac{\beta_0^2}{\epsilon^2}-\frac{\beta_1}{\epsilon}\right) 
-\left(\frac{\alpha_S}{\pi}\right)^3\left(\frac{\beta_0^3}{\epsilon^3}-\frac{2\beta_0 \beta_1}{\epsilon^2}+\frac{\beta_2}{\epsilon}\right) +\mathcal{O}(\alpha_S^4).\nonumber\\
\eea 
The coefficients at the various orders in the coupling constant $\beta_i$ are given by the QCD beta function~\cite{VanRitbergen:1997va,Czakon:2004bu,Baikov:2016tgj,Herzog:2017ohr}.

In order to obtain infrared finite cross sections we are required to perform a suitable redefinition of our parton distribution functions.	
\beq
f_i(x)=f_i^R\circ \Gamma,\hspace{1cm}(f\circ g )(z)=\int_0^1 dx dy f(x)g(y)\delta(xy-z).
\eeq
The infrared counter term $\Gamma$ consists of convolutions~\cite{H??schele2013}  of splitting functions $P_{ij}^{(n)}$~\cite{Moch2004,Vogt2004} and can be derived from the DGLAP equation. 
Its perturbative expansion required for an N$^3$LO accurate calculation of the differential Higgs boson production cross section is given by
\bea
\Gamma_{ij}&=&\delta_{ij}\delta(1-x)\nonumber\\
&+&\left(\frac{\alpha_S}{\pi}\right)\frac{P^{(0)}_{ij}}{\epsilon}\nonumber\\
&+&\left(\frac{\alpha_S}{\pi}\right)^2\left[\frac{1}{2\epsilon^2}\left(P^{(0)}_{ik}\circ P^{(0)}_{kj}-\beta_0  P^{(0)}_{ij}\right)+\frac{1}{2\epsilon}P^{(1)}_{kj}\right]\\
&+&\left(\frac{\alpha_S}{\pi}\right)^3\left[\frac{1}{6\epsilon^3}\left(P^{(0)}_{ik}\circ P^{(0)}_{kl}\circ P^{(0)}_{lj}-3\beta_0P^{(0)}_{ik}\circ P^{(0)}_{kj}+2\beta_0^2P^{(0)}_{ij}\right)\right.\nonumber\\
&&\left.+\frac{1}{6\epsilon^2}\left(P^{(1)}_{ik}\circ P^{(0)}_{kj}+2P^{(0)}_{ik}\circ P^{(1)}_{kj}-2\beta_0 P^{(1)}_{ij}-2 \beta_1 P^{(0)}_{ij}\right)+\frac{1}{3\epsilon}P^{(2)}_{ij}\right].\nonumber
\eea

In the effective theory with $n_f$ light flavours and the top quark decoupled from 
the running of the strong coupling constant,
the $\overline{\textrm{MS}}$-scheme Wilson coefficient 
 reads~\cite{Chetyrkin:1997un,Schroder:2005hy,Chetyrkin:2005ia,Kramer:1996iq}
\begin{eqnarray}
C(\mu^2) & = & - \frac{\alpha_S}{3 \pi v} \Bigg\{ 1 +\left(\frac{\alpha_S}{\pi} \right)\,\frac{11}{4} 
+ \left(\frac{\alpha_S}{\pi} \right)^2 \left[\frac{2777}{288} - \frac{19}{16} \log\left(\frac{m_t^2}{\mu^2}\right) -
 n_f\left(\frac{67}{96}+\frac{1}{3}\log\left(\frac{m_t^2}{\mu^2}\right)\right)\right] 
 \nonumber \\
  &&+ \left(\frac{\alpha_S}{\pi} \right)^3\Bigg[-\left(\frac{6865}{31104} 
  + \frac{77}{1728} \log\left(\frac{m_t^2}{\mu^2}\right) 
  + \frac{1}{18}\log^2\left(\frac{m_t^2}{\mu^2}\right)\right) n_f^2 \\
  && + \left(\frac{23}{32} \log^2\left(\frac{m_t^2}{\mu^2}\right) - 
  \frac{55}{54} \log\left(\frac{m_t^2}{\mu^2}\right)+\frac{40291}{20736} 
 - \frac{110779}{13824} \zeta_3 \right) n_f \nonumber\\
  &&  -\frac{2892659}{41472}+\frac{897943}{9216}\zeta_3 
  + \frac{209}{64} \log^2\left(\frac{m_t^2}{\mu^2}\right)
  - \frac{1733}{288}\log\left(\frac{m_t^2}{\mu^2}\right)\Bigg] +\mathcal{O}(\alpha_S^4) \Bigg\} \, .\nonumber
\end{eqnarray}
\bibliography{biblio}
\bibliographystyle{JHEP}

\end{document}